\documentstyle[preprint,eqsecnum,aps]{revtex}
\begin{document}
\draft

\title{
Meson-exchange Model for $\pi N$ Scattering and $\gamma N \rightarrow \pi N$ 
Reaction}

\author{T. Sato$^{1,2}$ and T.-S. H. Lee$^1$}
\address{
$^1$\underline{Physics Division, Argonne National Laboratory, Argonne, IL 60439-4843 USA} \\
$^2$\underline{Department of Physics, Osaka University, Toyonaka, Osaka 560, Japan}}

\maketitle

\begin{abstract}
An effective Hamiltonian consisting
of bare $\Delta \leftrightarrow\pi N$, $\gamma N$ vertex interactions and
energy-independent meson-exchange $\pi N \leftrightarrow \pi N, \gamma N$
transition operators is derived by applying
a unitary transformation to a model Lagrangian
with $N,\Delta,\pi$, $\rho$, $\omega$, and $\gamma$ fields.
With appropraite phenomenological form factors and coupling constants for
$\rho$ and $\Delta$, the model can give a good
description of $\pi N$ scattering phase shifts up to the $\Delta$ 
excitation energy region. 
It is shown that the best reproduction of the recent LEGS data of 
the photon-asymmetry ratios in $\gamma p \rightarrow \pi ^0 p$ reactions 
provides rather restricted
constraints on the coupling strengths 
$G_E$ of the electric 
$E2$ and $G_M$ of the magnetic $M1$ transitions of the bare 
$\Delta \leftrightarrow \gamma N$ vertex and the less 
well-determined coupling constant $g_{\omega NN}$ of $\omega$ meson. 
Within the ranges that $G_M = 1.9 \pm 0.05$,  
$G_E = 0.0 \pm 0.025$, and $7 \leq g_{\omega NN}\leq 10.5$,
the predicted differential cross
sections and photon-asymmetry ratios are in an overall good agreement with the
data of $\gamma p \rightarrow \pi ^0 p$, $\gamma p \rightarrow \pi ^+ n$,
and $\gamma n\rightarrow \pi ^- p$ reactions from 180 MeV to the $\Delta$ excitation region.  
The predicted $M_{1^+}$ and $E_{1^+}$ multipole amplitudes are also in good
agreement with the empirical values determined by the amplitude analyses.  The constructed
effective Hamiltonian is free of the nucleon renormlization problem and hence is
suitable for nuclear many-body calculations.  We have also shown that the
assumptions made in the $K$-matrix method, commonly used in extracting empirically the
$\gamma N \rightarrow \Delta$ transition amplitudes from the data, are consistent with
our meson-exchange dynamical model.  It is found that the helicity amplitudes
calculated from our bare $\gamma N \rightarrow \Delta$ vertex are in good agreement
with the predictions of the constituent quark model.  The differences between these
bare amplitudes and the dressed amplitudes, which are closer to the empirical values
listed by the Particle Data Group, 
are shown to be due to the non-resonant meson-exchange mechanisms.  
Within the range $ 7 \le g_{\omega NN} \le 10.5$ of the $\omega$ meson coupling
favored by the data of
the photon-asymmetry ratios in $\gamma p \rightarrow \pi^0 p$ reactions, 
our values of the $E2/M1$ ratio for the 
$\gamma N \rightarrow \Delta$ transition are
(0.0 $\pm$ 1.3)\% for the bare vertex and (-1.8 $\pm$ 0.9)\% for the dressed vertex.

\bigskip
\bigskip
\bigskip

\end{abstract}

\newpage

\section{Introduction}
The main objective of investigating photo- and electro-production of mesons on the
nucleon is to study the structure of the nucleon excited states (N$^*$).
This has been pursued actively \cite{donnachie} during the period around 1970.
With the developments at several electron facilities since 1980, more extensive
investigations of the $\Delta$-excitation have been carried out both experimentally
and theoretically \cite{drechel}. Apart from the need of precise and extensive measurements 
which will soon be possible at CEBAF and Mainz, an accurate understanding of 
the $N^*$ structure can be obtained only when an
appropriate reaction theory is developed to separate the reaction mechanisms from
the hadron structure in the $\gamma N\rightarrow \pi N,\pi\pi N$ reactions.
The importance of this theoretical effort can be understood by recalling many years
experiences in the development
of nuclear physics.  For example, the information about the deformation of
$^{12}C$ can be extracted from $^{12}C(p,p')^{12}C^* (2^+ ,2.44$ MeV) inelastic scattering
only when a reliable reaction theory \cite{feshbach}, such as the Distorted-wave 
Impulse Approximation or the coupled-channel method, is used to calculate the
initial and final proton-$^{12}C$ interactions.  Accordingly, one expects that 
the $N^*$ structure
can be determined only when the interactions in its decay channels $\gamma N$,
$\pi N$, and $\pi\pi N$ can be calculated from a reliable reaction theory.
It is the objective of this work to address this problem from the point of view of
meson-exchange models.  In contrast to approaches based on the dispersion
relations \cite{donnachie} or the $K$-matrix method \cite{olsson,davidson91,davidson90,li}, 
our approach is aimed at not
only an investigation of the $N^*$ structure but also on the application of the
constructed model to a consistent calculation of $N^*$ in nuclear many-body systems.

The meson-exchange models have been very successful in describing nucleon-nucleon
interactions \cite{machleidt}, electroweak interaction
currents \cite{riska,rho}, meson-meson scattering \cite{julich1}, and
meson-nucleon scattering \cite{pearce,yang1,julich2}. It is therefore reasonable
to expect that the same success can also be achieved in the investigation of pion
photo- and electro-production. This possibility has, however, not been fully 
explored. The dynamical models of pion photoproduction developed
in Refs.\ \cite{tanabe,yang2,nbl} did contain the well-established
meson-exchange mechanisms of pion photoproduction, but phenomenological
separable potentials were used to describe the $\pi N$ multiple 
scattering.  The improvement made in Ref.\ \cite{leepearce} suffered from the theoretical
inconsistency in defining the meson-exchange $\pi N$ interaction and $\pi N \rightarrow
\gamma N$ transition. 
The model developed in Refs.\ \cite{surya1,surya2} also does not treat
meson-exchange completely since a zero-range contact term is introduced to replace
the particle-exchange terms of their $\pi N$ potential.
In all of these models, the  incomplete treatment of the meson-exchange interactions
leads to some uncertainties
in interpreting the parameters characterizing the $\gamma N \rightarrow \Delta$
vertex which is the main interest in testing hadron models.
The formulation developed in Ref.\ \cite{araki} can, in principle, be
used to examine the meson-exchange mechanisms in pion photoproduction, but has
not been pursued numerically.
In this work, we will try to improve the situation by 
appling the unitary transformation method developed in Ref.\ \cite{sato4}
to derive from a model Lagrangian 
an effective Hamiltonian for a consistent meson-exchange description of
both the $\pi N$ scattering and pion photoproduction. 
Furthermore, the constructed model can be directly used to improve and extend the
$\pi NN$ Hamiltonian developed in Ref.\ \cite{lee86} to also describe the electromagnetic
$\Delta$ excitation in intermediate energy nuclear reactions.

The starting point of constructing a meson-exchange model is
a model Lagrangian of relativistic quantum field theory. The form of the
Lagrangian is constrained by the observed symmetries of fundemental 
interactions, such as Lorentz invariance, isospin conservation, chiral 
symmetry, and gauge invariance. The most common approach \cite{klein} is
to find an appropriate three-dimensional reduction of the Ladder Bethe-Salpeter 
equation of the considered model Lagrangian.  
The meson-exchange potentials are then identified with the driving terms of
the resulting three-dimensional scattering equation.
The most recent examples are the $\pi N$ models
developed in Refs.\ \cite{pearce,yang1,julich2,surya1}. The extension of this 
approach to investigate pion photoproduction has also been made in Ref.\ \cite{surya2}.

Alternatively, one can construct a meson-exchange model by deriving
an effective Hamiltonian from the considered model Lagrangian. 
Historically, two approaches have been developed. The first one is to 
use the Tamm-Dancoff approximation \cite{tamm}. This method leads
to an effective Hamiltonian which is energy-dependent and contains unlinked
terms, and hence can not be easily used in nuclear many-body calculations.  
A more tractable approach is to apply the method of unitary transformation 
which was developed by 
Fukuda, Sawada and Taketani \cite{fuku}, and independly by Okubo\cite{okubo}. 
This approach, called the FST-Okubo method, 
has been very useful in investigating nuclear electromagnetic
currents \cite{ohtawaka,gari,sato1} and relativistic descriptions of
nuclear interactions \cite{glockle1,glockle2,sato2,sato3}.
The advantage of this approach is that
the resulting effective Hamiltonian 
is energy independent and can readily be used in nuclear many-body calculation.
Motivated by the investigation of the $\pi NN$ dynamics \cite{lee86,garcilazo}, 
this method has been
extended in Ref.\ \cite{sato4} to derive an effective theory
involving pion production channels. In this work, we adopt this method 
to develop a dynamical model for $\pi N$ scattering and $\gamma N
\rightarrow \pi N$ reactions.

It is necessary to explain here how our approach is related to the approach based
on the Chiral Perturbation Theory (CHPT) \cite{bernard}.  Since chiral symmetry is a 
well-established dynamical symmetry of strong interactions, it should be used to 
constrain our starting Lagrangian.  This leads us to assume that our starting Lagrangian 
is an effective Lagrangian for generating the tree-diagrams in CHPT.
The parameters are then completely determined by the well-established chiral dynamics
such as PCAC and current algebra.  Therefore, our model and CHPT 
are identical in leading orders.  The differences come from how the unitarity is implemented
to account for the $\pi N$ multiple scattering.  In the spirit of CHPT,
 the ``low" momentum pions are considered as weakly interacting Goldstone
bosons and hence their interactions with the nucleon can be treated as 
perturbations \cite{weinberg}.  This amounts to restoring the unitarity perturbatively 
by calculating
loop corrections order by order.  It is then necessary to include more terms in the
effective Lagrangian.  A phenomenological procedure is then unavoidable to
determine the accompanied low-energy constants.

In the meson-exchange model, one hopes to describe the $\pi N$ multiple scattering in the
entire kinematic region including the highly non-perturbative $\Delta$ excitation
region.  The essential
assumption is that the $\pi N$ multiple scattering is governed by a few-body Schrodinger
equation with the driving terms calculated from the starting Lagrangian 
in a perturbation expansion in the coupling constants.
This can be realized in practice only when the driving terms are regularized
by appropriate phenomenological form factors.  Qualitatively speaking, the meson-exchange
model is an alternative to CHPT in the kinematic region where perturbative
calculations become very difficult or impossible.  
Both approaches involve phenomenological
parameters.  The success of each approach depends on whether these 
parameters can be interpreted theoretically.

In this work we will focus on the $\Delta$ excitation and will limit our
investigation to the energy region where $2\pi$ production is negligibly small.
By applying the unitary transformation of Ref.\ \cite{sato4} to a model Lagrangian
for $N$, $\Delta$, $\pi$, $\rho$, $\omega$, and $\gamma$ fields, 
we have obtained an
effective Hamiltonian consisting of bare $\Delta \leftrightarrow \pi N, \gamma N$
vertex interactions and energy-independent $\pi N \leftrightarrow \pi N, \gamma N$
transition operators.  The $\pi N$ scattering phase
shifts \cite{arndt,hohler,koch} are used to determine the hadronic part of the constructed
effective Hamiltonian which has only seven parameters for defining the vertices of
the meson-exchange $\pi N$ potential and the $\Delta \leftrightarrow \pi N$ transition.
The strong vertex functions in the $\gamma N \rightarrow \pi N$ transition operator are
then also fixed.  This is a significant improvement over the previous dynamical models
\cite{tanabe,yang2,nbl} in which the employed separable potentials have no dynamical
relation with the pion photoproduction operator.  A consistent description of the
$\pi N$ scattering and $\gamma N \rightarrow \pi N$ transition is crucial for
separating the reaction mechanisms due to meson-exchange non-resonant interactions from
the total $\gamma N \rightarrow \Delta$ transition.

Once the hadronic part of the effective Hamiltonian is determined, the resulting pion
photoproduction amplitude has only three adjustable parameters:
$G_M$ of magnetic $M1$ and $G_E$ of electric $E2$ transitions of 
the bare $\gamma N\rightarrow \Delta$ vertex, and the less
well-determeined $\omega NN$ coupling constant.  We will determine these three parameters
by considering the most recent LEGS data \cite{legs}
of the photon-asymmetry ratios in $\gamma p \rightarrow \pi^0 p$
reactions.  The resulting parameters are then tested against very extensive
data in Refs.\ \cite{sandorfi,menze,dugan}.

It is customary to test hadron models by comparing the theoretical predictions 
of $N^* \rightarrow \gamma N$ transition amplitudes with
the empirical values listed by the Particle Data Group [PDG] \cite{particle}.  
Since the first
systematic calculation \cite{isgur} based on the constituent quark model was performed, it
has been observed that the predicted $\Delta \rightarrow \gamma N$ transition amplitudies
\cite{isgur,capstick,bijker,close,keister} are significantly smaller than the empirical values listed by
PDG \cite{particle}.  While the problem may be due to the limitations of the constituent quark
model, it is necessary to recognize
that the empirical values of PDG are obtained by using the $K$-matrix method
\cite{olsson,davidson91,davidson90,li} or dispersion relation \cite{donnachie}.
Both approaches contain assumptions about the non-resonant contributions to the
$\gamma N \rightarrow \Delta$ transition and must be justified from a dynamical
point of view.  Within our dynamical model, we will address this point concerning the
$K$-matrix method.  This leads us to identify our bare $\gamma N\rightarrow\Delta$
vertex with the constituent quark model.  The dispersion relation approach
\cite{donnachie,tiator} is defined
in a very different theoretical framework and therefore is beyond the scope of this
investigation.

In section II, we will use a simple model Lagragian to
explain how an effective Hamiltonian can be constructed by using  
the unitary transformation method of Ref.\ \cite{sato4}.
The method is then applied to realistic Lagrangians to derive in sections III and IV
an effective Hamiltonian for $\pi N$ scattering and  pion photoproduction.
The equations for calculating the $\pi N$ scattering and $\gamma N \rightarrow \pi N$
amplitudes are also presented there.  The relationship with the $K$-matrix method
are then established.
Results and discussions are given in section V.  
The conclusions and discussions of future studies are given in section VI.

\section{Method of Unitary Transformation}

To explain the unitary transformation method of Ref.\ \cite{sato4} (will be referred to
as the SKO method), 
it is sufficient to consider a simple system consisting of only 
neutral pions and fictituous $\sigma$ mesons. The objective is to derive an effective 
Hamiltonian from the following Lagrangian density
\begin{equation}
L(x) = L_0(x) + L_I(x), \nonumber
\end{equation}
where $L_0(x)$ is the usual noninteracting Lagrangian, and the interaction term is taken to be
\begin{equation}
L_I(x) = - g_{\sigma\pi\pi} \phi_{\pi}^2(x)\phi_{\sigma}(x).
\end{equation}
The Hamiltonian can be derived from Eq.\ (2.1) by using the standard method of canonical
quantization. In the second-quantization form, we obtain (in the convention of
Bjorkin and Drell \cite{bj})
\begin{equation}
H  =  H_0 + H_I ,
\end{equation}
with
\begin{eqnarray}
H_0 & = & \int d\vec{k}
 (  E_{\pi}(k)a^{\dagger}(\vec{k})a(\vec{k}) +
 E_{\sigma}(k) b^{\dagger}(\vec{k})b(\vec{k})), \\
H_I  &  = & g_{\sigma\pi\pi}\int\frac{d\vec{k}_1d\vec{k}_2d\vec{k}}
       {\sqrt{(2\pi)^3
           8E_{\pi}(k_1)E_{\pi}(k_2)E_{\sigma}(k)}}  \nonumber  \\
&\times& ( [2a^{\dagger}(\vec{k}_1)a(\vec{k}_2)
b(\vec{k})\delta(\vec{k}_1-\vec{k}_2-\vec{k})
 + a^{\dagger}(\vec{k}_1)a^{\dagger}(\vec{k}_2)b(\vec{k})
\delta(\vec{k}_1+\vec{k}_2-\vec{k}) \nonumber  \\
&+& a(\vec{k}_1)a(\vec{k}_2)b(\vec{k})
\delta(\vec{k}_1+\vec{k}_2+\vec{k}) ] + [\mbox{h.c.}] ),
\end{eqnarray}
where $a^{\dagger}(\vec{k})$ and $b^{\dagger}(\vec{k})$ are, respectively, the
creation operators for $\pi$ and $\sigma$ particles, 
$E_{\alpha}(k) = \sqrt{m_{\alpha}^2 + k^2}$ is the free energy for the
particle $\alpha$, and $[h.c.]$ means taking the hermitian conjugate of the first
term in the equation. We further assume that the mass of the $\sigma$ meson
is heavier than two-pion mass; i.e.  $m_{\sigma} > 2 m_{\pi}$.

Because of the intrinsic many-body problem associated with the starting quantum
field theory, it is not possible to solve exactly the equation of motion 
for meson-meson scattering defined by the above Hamiltonian.
A simplification is
obtained by assuming that in the low and intermediate energy regions, only
"few-body" states are active and must be treated explicitly.
The effects due to "many-body" states are absorbed in effective interaction
operators which can be calculated in a perturbation expansion in coupling
constants.  This few-body approach to field theory was pioneered by
Amado \cite{aaron}.  
In the SKO approach, this is achieved by first decomposing the
interaction Hamiltonian $H_I$ Eq.\ (2.5) into two parts
\begin{eqnarray}
H_I &= &H_I^P + H_I^Q\\
H_I^P & = & g_{\sigma\pi\pi}\int \frac{d\vec{k}_1d\vec{k}_2d\vec{k}}
       {\sqrt{(2\pi)^3
           8E_{\pi}(k_1)E_{\pi}(k_2)E_{\sigma}(k)}}
(a^{\dagger}(\vec{k}_1)a^{\dagger}(\vec{k}_2)b(\vec{k})
\delta(\vec{k}_1+\vec{k}_2-\vec{k})  + [\mbox{h.c.}])  \label{hip} \\
H_I^Q & = & g_{\sigma\pi\pi}\int \frac{d\vec{k}_1d\vec{k}_2d\vec{k}}
       {\sqrt{(2\pi)^3
         8E_{\pi}(k_1)E_{\pi}(k_2)E_{\sigma}(k)}}
 \nonumber  \\
 & ( &
[ 2a^{\dagger}(\vec{k}_1)a(\vec{k}_2)b(\vec{k})\delta(\vec{k}_1-\vec{k}_2-\vec{k})
  + a(\vec{k}_1)a(\vec{k}_2)b(\vec{k})
\delta(\vec{k}_1+\vec{k}_2+\vec{k}) ] + [\mbox{h.c.}])
\end{eqnarray}
The elementary processes induced by $H_I^P$ are illustrated in the upper half of
Fig.\ 1.
For $m_{\sigma} > 2m_{\pi}$, the $\sigma\rightarrow \pi\pi$ decay and  
$\pi\pi \rightarrow \sigma$ annihilation are 'real processes' and 
can take place in free space. On the other hand, the processes $\pi 
\leftrightarrow \pi\sigma$ and vacumm $\leftrightarrow \pi\pi\sigma$ 
induced by $H_I^Q$ are 'virtual processes'(lower part of Fig.\ 1). They 
can not occur in free space 
because of the energy-momentum conservation.
The essence of the SKO method is
to systematically eliminate the virtual processes 
from the considered Hamiltonian by using unitary transformations. 
As a result the effects  of 'virtual proceesses' are included as
effective operators in the transformed Hamiltonian.

The transformed Hamiltonian is defined as

\begin{eqnarray}
H' &=& UHU^+ \\
\nonumber
&=& U(H_0 + H^P_I + H^Q_I ) U^+ \,,
\end{eqnarray}
where $U = exp(-iS)$ is a unitary operator defined by a hermitian operator S.\,\, By expanding
$U = 1-iS + ...$\,, the transformed Hamiltonian can be written as

\begin{eqnarray}
H' &=& H_0 + H^{P}_I + H^{Q}_I + [H_0 , iS\,] \\
\nonumber
&+& [H_I , iS\,] + {1\over 2!} \, \Big[ [H_0 , iS\,] , iS\,\Big] + \cdots
\end{eqnarray}
To eliminate from Eq.\ (2.10) the virtual processes which are of first-order in 
the coupling constant $g_{\sigma\pi\pi}$, the SKO method imposes the condition
that

\begin{equation}
H^{Q}_{I} + [H_0 , iS\,] = 0 \,.
\end{equation}
Since $H_0$ is a diagonal operator in Fock-space , 
Eq.\ (2.11) clearly implies that $iS$ must have the same
operator structure of $H^{Q}_{I}$.  

To simplify the presentation, we write $H^Q$ as
\begin{equation}
H^Q_I = \sum_{n} \int F_n \, O_n \, d\vec k_1 d\vec k_2 \, d\vec k \,,
\end{equation}
where $O_n$ denotes the part 
containing creation and annihilation operators, and $F_n$ is
the rest of $nth$ term in Eq.\ (2.8).  
In the form of Eq.\ (2.12), the solution of Eq.\ (2.11) can be written as

\begin{equation}
iS = \sum_{n} i \int S_n \, O_n \, d\vec k_1 d\vec k_2 \, d\vec k \,.
\end{equation}
Our task is to find $S_n$ by solving Eq.\ (2.11).  Considering two eigenstates
$|i>$ and $|f>$  
of the free Hamiltonian $H_0$ 
such that $<f|O_n |i> = 1$, Eqs.\ (2.11), (2.12) and (2.13) then lead to
\begin{equation}
iS_n = {-1\over E_f - E_i}\, F_n  \,.
\end{equation}
Note that $E_{i}$ and $E_f$ are the eigenvalues of free Hamiltonian $H_0$, and hence
the solution $S_n$ is independent of 
the collision energy $E$ of the total Hamiltonian $H$. This is an important
feature distingushing our approach from the Tamm-Dancoff approximation.
By using the above relation, 
it is easy to verify that the solution of 
the operator equation (2.11) is 
\begin{eqnarray}
iS & = & g_{\sigma\pi\pi} \int \frac{d\vec{k}_1d\vec{k}_2d\vec{k}}
       {\sqrt{(2\pi)^3
           8E_{\pi}(k_1)E_{\pi}(k_2)E_{\sigma}(k)}}
 \nonumber  \\
 & (&
[ 2a^{\dagger}(\vec{k}_1)a(\vec{k}_2) b(\vec{k})
\frac{\delta(\vec{k}_1-\vec{k}_2-\vec{k})}
     {-E_{\pi}(k_1)+E_{\pi}(k_2)+E_{\sigma}(k)}
\nonumber \\
  &+ & a(\vec{k}_1)a(\vec{k}_2)b(\vec{k})
\frac{\delta(\vec{k}_1+\vec{k}_2+\vec{k})}
     {E_{\pi}(k_1)+E_{\pi}(k_2)+E_{\sigma}(k)}     
 ] + \mbox{h.c.} ) \,.
\end{eqnarray}

By using Eq.\ (2.11), Eq.\ (2.10) can be written as
\begin{eqnarray}
H^\prime = H_0 + H_I^\prime \,, 
\end{eqnarray}
with
\begin{eqnarray}
H_I^\prime = H_I^P + [H_I^P,iS\,] + \frac{1}{2} \,[H_I^Q,iS\,]
+ \mbox{higher order terms}\,.
\end{eqnarray}
Since $H_I^P$, $H_I^Q$, and $S$ are all of the first order in the coupling constant
$g_{\sigma\pi\pi}$, all processes included in the second and third terms of
the $H_I^\prime$ are of the order of $g_{\sigma\pi\pi}^2$. But some of them
are 'real processes', such as the $\sigma$-exchange $\pi\pi$ interaction 
and $\pi$-exchange $\pi\pi\rightarrow\sigma\sigma$ transition
, as illustrated in the upper half of Fig.\ 2.
The other processes are 'virtual processes'. An example is 
the emission of two $\sigma$ mesons by a pion 
illustrated in the lower half of Fig.\ 2.
We therefore rewrite Eq.\ (2.17) as
\begin{equation}
H_I'  =  H_I^P + H_I'^P + H_I'^Q + \sum_{n \geq 3} O(g_{\sigma\pi\pi}^n) \,,
\end{equation}
where
\begin{eqnarray}
H_I'^P &=& ([H_I^P,iS] +\frac{1}{2}[H_I^Q,iS])^P \,, \\
H_I'^Q  &=& ([H_I^P,iS] + \frac{1}{2}[H_I^Q,iS])^Q \,.
\end{eqnarray}
In the above two definitions, $H_I'^P(H_I'^Q)$ is obtained by evaluating
the commutators using Eqs.\ (2.7), (2.8), and (2.15) and keep only the
real (virtual) processes in the results.

The next step is to perform a second unitary tranformation to eliminate $H^{'Q}$.
In this paper we only consider the effective Hamiltonian up to second-order in 
the coupling constant, so we do not need to consider the second unitary transformation.  The effective
Hamiltonian is then obtained by dropping $H'^Q$ and higher-order terms in
Eq.\ (2.18)
\begin{equation}
H_{eff} = H + H^{P}_{I} + H^{'P}_{I} \,,
\end{equation}
where $H_I^P$ is defined by Eq.\ (2.7), and $H_I'^P$ can be calculated from
Eq.\ (2.19) by using the solution Eq.\ (2.15) for $S$.
Note that $H_I'^P$ contains a $\sigma-\pi$-loop correction to the pion mass term.
If we choose the pion mass in $H_0$ as the physical mass, this loop correction should be
dropped from $H_I'^P$ in order to avoid double counting. 
A similar situation will be encountered in the derivation of $\pi N$ effective 
Hamiltonian. In our model we choose the physical masses for $N$ 
and $\pi$ in the free Hamiltonian and hence the loop corrections in the
effective Hamiltonian are also dropped.  This phenomenological procedure saves
us from facing the complicated mass renormalization problem in solving
scattering problems.

Because of Eq.\ (2.14), Eqs.\ (2.12) and (2.13) lead to the
following simple relation

\begin{equation}
< f\,|iS\,|i > \, = {-1\over E_f - E_i}\, <f\,|H^Q_{I}\,|i > \,.
\end{equation}
where $|f>$ and $|i>$ are two eigenstates of $H_0$.
With the above relation, the calculation of $H^{'P}_I$  Eq.\ (2.19) becomes

\begin{eqnarray}
<f\,|H^{'P}_I\,|i> &=& \sum_{n} \Bigg\{ <f\,|H^P_I\,| n>\,< n\,|H^Q_I\,|i >\,
{1\over E_i - E_n} \nonumber \\
\nonumber
 &-& <f\,| H^Q_I\,|n> \, <n\,|H^P_I\,|i> {1\over E_n - E_f} \\
&+& <f\,|H^Q_I\,|n >\,<n\,|H^Q_I\,|i > {1\over 2}\, \Bigg[{1\over E_i - E_n} -
{1\over E_n - E_f} \Bigg] \Bigg\} \,.
\end{eqnarray}
The calculation of the matrix element of $H^{'P}_{I}$ therefore has a very simple
rule.  For a given choice of basis states 
$|i>$ and $|f>$, the allowed intermediate state $n$ is
determined by the operator structure of $O_n$ in Eqs.\ (2.12) and (2.13).  The
denominator in Eq.\ (2.23) can easily be written down by using eigenvalues
of the free Hamiltonian $H_0$.  

Evaluating $H_I^P$ and $H_I'^P$ explicitly within the
coupled $\pi\pi\oplus\sigma$ sapce, 
Eq.\ (2.21) can be cast into 
the following more familiar form for $\pi\pi$ scattering
\begin{equation}
H_{eff}^{\pi\pi} = H_0 + f_{\sigma\leftrightarrow \pi\pi}
+ V_{\pi\pi}.
\end{equation}
Here $H_0$ is the free Hamiltonian operator for $\pi$ and $\sigma$ mesons.
The second term describes the $\sigma \leftrightarrow \pi\pi$ transition with the
following matrix element
\begin{eqnarray}
<\vec{k}_1\vec{k}_2 \mid f_{\pi\pi,\sigma}\mid\vec{k}>
=\frac{\sqrt{2}g_{\sigma\pi\pi}}{\sqrt{(2\pi)^3 8E_{\pi}(k_1)
E_{\pi}(k_2)E_{\sigma}(k)}} \,.
\end{eqnarray}
The $\pi\pi$ potential $V_{\pi\pi}$ is obtained by using Eq.\ (2.23) to 
calculate $H_I'^P$ between two $\pi\pi$ states. 
For $|i>=|\vec{k}_{i1}\vec{k}_{i2} > =\sqrt{\frac{1}{2}}
[a^+_{\vec{k}_{i1}}a^+_{\vec{k}_{i2}}] |0>$ 
and $<f| = <\vec{k}_{f1}\vec{k}_{f2}| 
= <0|[a_{\vec{k}_{f1}}a_{\vec{k}_{f2}}]\sqrt{\frac{1}{2}}$
the possible intermediate states in Eq.\ (2.23) are 
$|\pi\pi\sigma>$ and $|\pi\pi\pi\pi\sigma>$ states.
Inserting these intermediate states into Eq.\ (2.23) and 
carring out straightforward
operator algebra, we obtain
\begin{eqnarray}
V_{\pi\pi} = V_{\pi\pi}^s + V_{\pi\pi}^t
\end{eqnarray}
with the following the matrix elements between two $\pi\pi$ states
\begin{eqnarray}
<\vec{k}_{f1} \vec{k}_{f2} \mid V_{\pi\pi}^s\mid \vec{k}_{i1} \vec{k}_{i2} >
=  \frac{g_{\sigma\pi\pi}^2}{(2\pi)^3}
\frac{1}{\sqrt{2E_{\pi}(k_{f1})}}\frac{1}{\sqrt{2E_{\pi}(k_{f2})}}
\frac{1}{\sqrt{2E_{\pi}(k_{i1})}}\frac{1}{\sqrt{2E_{\pi}(k_{i2})}} 
\nonumber \\
\times [D^{(-)}_\sigma(k_{i1}+k_{i2})+D^{(-)}_\sigma(k_{f1}+k_{f2})] \,,
\end{eqnarray}

\begin{eqnarray}
<\vec{k}_{f1}\vec{k}_{f2}\mid V_{\pi\pi}^t \mid \vec{k}_{i1}\vec{k}_{i2}>
= \frac{g_{\sigma\pi\pi}^2}{(2\pi)^3}
\frac{1}{\sqrt{2E_{\pi}(k_{f1})}}\frac{1}{\sqrt{2E_{\pi}(k_{f2})}}
\frac{1}{\sqrt{2E_{\pi}(k_{i1})}}\frac{1}{\sqrt{2E_{\pi}(k_{i2})}}
\nonumber \\
\times  [D_\sigma(k_{i1}-k_{f2})+D_\sigma(k_{i2}-k_{f2}) 
          +D_\sigma(k_{i1}-k_{f1})+D_\sigma(k_{i2}-k_{f1})] \,,
\end{eqnarray}
where 
\begin{eqnarray}
D_\sigma(k) &=&\frac{1}{k^2-m_\sigma^2}=D^{(+)}(k)+D^{(-)}(k) \,,
\end{eqnarray}
with
\begin{eqnarray}
D_\sigma^{(\pm)}(k)&=&\frac{1}{2E_\sigma(k)}
\frac{\pm 1}{k_0\mp E_\sigma(k)} \,.
\end{eqnarray}

This completes the illustration of the SKO method in deriving an effective Hamiltonian
from a model Lagrangian of relativistic quantum field theory. The extension of
the method to consider more realistic Lagrangians is straightforward and
will not be further detailed. In the following sections, we will simply write
down the starting Lagrangians and the resulting effective Hamiltonians up 
to second order in the coupling constants for
$\pi N$ scattering and the $\gamma N\rightarrow \pi N$ reaction.

\section{$\pi$-N Scattering}

We start with the following commonly assumed\cite{nbl} Lagrangian 
for $N,\Delta,\pi$ and $\rho$ fields
\begin{eqnarray}
L(x)=L_0(x)+L_I(x),
\end{eqnarray}
where $L_0(x)$ is the usual noninteracting Lagrangian, and the
interaction is taken to be
\begin{equation}
L_I(x) = L_{\pi NN}(x) + L_{\pi N\Delta}(x)+ L_{\rho NN}(x) 
+ L_{\rho\pi\pi}(x), 
\end{equation}
with( in the convention of Bjorkin and Drell \cite{bj})
\begin{eqnarray}
L_{\pi NN}(x)& = &  -\frac{f_{\pi NN}}{m_{\pi}}\bar{\psi}_N(x)
  \gamma_5\gamma_{\mu}\vec{\tau}\psi_N(x)\partial^{\mu}
\cdot \vec{\phi}_{\pi}(x), \\
L_{\pi N\Delta}& =& 
 \frac{f_{\pi N\Delta}}{m_{\pi}}\bar{\psi}_{\Delta}^{\mu}(x)
  \vec{T}\psi_N(x)\cdot\partial_{\mu}\vec{\phi}_{\pi}(x) + [\mbox{h.c.}], \\
L_{\rho NN}(x) &=& 
 g_{\rho NN}\bar{\psi}_N(x) \frac{\vec{\tau}}{2}\cdot
 [ \gamma_{\mu}\vec{\phi}^{\mu}_{\rho}(x)
- \frac{\kappa_{\rho}}{2m_{N}}\sigma_{\mu\nu}\partial^{\nu} 
  \vec{\phi}_{\rho}^{\mu}(x)] \psi_N(x), \\
L_{\rho\pi\pi}(x)&=&g_{\rho\pi\pi}
(\vec{\phi}_{\pi}\times\partial_{\mu}\vec{\phi}_{\pi})
\cdot\vec{\phi}_{\rho}^{\mu}.
\end{eqnarray}
Here $\vec{T}$ is a $N\rightarrow \Delta$ isospin transition operator defined
by the reduced matrix element\cite{edmond} $<\frac{3}{2} ||T ||\frac{1}{2} > = -
<\frac{1}{2} || T^\dagger || \frac{3}{2} > = 2$.
By using the standard canonical quantization, a Hamiltonian can be derived 
from the above Lagrangian except the term involving the 
$\Delta$ field. The difficulty of quantizing the $\Delta$ field
is well known, as discussed, for example, in textbook \cite{lourie}
and Ref.\ \cite{benm}. 
As part of our phenomenology, we take the simplest prescription by 
imposing the following
anti-commutation relation 
\begin{eqnarray}
\big\{ \Delta_{\vec{p}}, \Delta^+_{\vec{p}^\prime}\big\} 
= \delta(\vec{p}-\vec{p}^\prime),
\end{eqnarray}
where $\Delta_{\vec{p}}$($\Delta^+_{\vec{p}}$) is the annihilation(creation)
operator for a $\Delta$ state.
This choice then leads\cite{lourie} 
to the $\Delta$ propagator given later in Eq.\ (3.18).
The alternative approaches proposed in Ref.\ \cite{benm} will not be considered.
 
Following the procedure described in section II, the next step is to 
decompose the resulting Hamiltonian into a $H^P_I$ for 
'physical processes' and a $H_I^Q$ for 'virtual processes'. 
 From the expressions Eqs.\ (3.3)-(3.6), it is clear
that the real processes in this case are $\Delta \leftrightarrow \pi N$ and $\rho
\leftrightarrow \pi\pi$ transitions which can take
place in free space(because $m_\Delta > m_N + m_\pi$ and $m_\rho > 2 m_\pi$).
The virtual processes are $N \leftrightarrow \pi N$,
$N \leftrightarrow \rho N$ ,$ N \leftrightarrow \pi \Delta$, and
$\pi \leftrightarrow \pi \rho$ transitions. These virtual processes can be eliminated
by introducing a unitary transformation operator $S$ which can be determined
by using the similar method in obtaining the solution
Eqs.\ (2.13)-(2.14). Here, we of course encounter a much more 
involved task 
to account for the Dirac spin structure, isospin, and also the
anti-particle components of $N$ and $\Delta$.
To see the main steps, we present in Appendix
A an explicit derivation of the potential due to the $L_{\pi NN}$ term Eq.\ (3.3). 

For practical applications, it is sufficient to 
present our results in the coupled $\pi N \oplus \Delta$ subspace in which the
$\pi N$ scattering problem will be solved. The resulting effective Hamiltonian
then takes the following form
\begin{eqnarray}
H_{eff}^{\pi N} = H_0 + \Gamma_{\Delta\leftrightarrow\pi N} + v_{\pi N},
\end{eqnarray}
where $H_0$ is the free Hamiltonian for $\pi,N$ and $\Delta$.
Note that $\Gamma_{\Delta\leftrightarrow\pi N}$ (Figs.\ 3a and 3b)
is the only vertex interaction in the constructed effective Hamiltonian.
Our model is therefore distinctively different from the previous meson-exchange 
$\pi N$ models\cite{pearce,yang1,julich2,surya1} 
which all involve a bare nucleon state $N_0$ and a $N_0\leftrightarrow\pi N$ 
vertex. 

The $\pi N$ potential $v_{\pi N}$ in Eq.\ (3.8) is found to be 
\begin{eqnarray}
v_{\pi N} = v_{N_D} + v_{N_E} + v_\rho + v_{\Delta_D} + v_{\Delta_E},
\end{eqnarray}
where $v_{N_D}$ is the direct nucleon pole term (Fig.\ 3c), $v_{N_E}$ the 
nucleon-exchange term (Fig.\ 3d), $v_{\rho}$ the $\rho-$exchange term (Fig.\ 3e),
$v_{\Delta_D}$ the interaction  due to the anti-$\Delta$ component of
the $\Delta$ propagation (Fig.\ 3f), and $v_{\Delta_E}$ 
the $\Delta$-exchange term (Fig.\ 3g). To simplify the presentation,
we will only give the matrix element of $v_{\pi N}$ in the $\pi N$ center of
mass frame. The initial and final four-momenta $k^{\mu}, k^{'\mu}$ for pions and
$p^{\mu}, p^{'\mu}$ for the nucleons in Fig.\ 3 are therefore defined as
\begin{eqnarray}
k^{\mu} &=& (E_{\pi}(k),\vec{k}), \nonumber \\
p^{\mu} &=& (E_{N}(k),-\vec{k}),\nonumber \\
k^{'\mu} &=& (E_{\pi}(k'),\vec{k}), \nonumber \\
p^{'\mu} &=& (E_{N}(k'), -\vec{k}).
\end{eqnarray}
In terms of these variables, the matrix element of each term of Eq.\ (3.9)
between two $\pi N$ states can be written as 
\begin{eqnarray}
<\vec{k}'i',m_s' m_\tau '|v_{\alpha}|\vec{k} i,m_s,m_\tau> &=&
 \frac{1}{(2\pi)^3}\frac{1}{\sqrt{2E_{\pi}(k')}}\sqrt{\frac{m_N}{E_N(k')}} 
\frac{1}{\sqrt{2E_{\pi}(k)}}\sqrt{\frac{m_N}{E_N(k)}}.
\nonumber \\
 &\times&  \bar{u}_{-\vec{k}',m_s',m_{\tau}'} 
I_{\alpha}(\vec{k}' i',\vec{k} i)u_{-\vec{k},m_s,m_{\tau}} \,, 
\end{eqnarray}
where $u_{\vec{p},m_s,m_\tau}$ is the Dirac spinior, 
$m_s$ and $m_\tau$ are the nucleon spin and isospin quantum
numbers, $i$ and $i'$ are the pion isospin components. The interaction mechanisms
are contained in the functions $I_{\alpha}(\vec{k}' i',\vec{k} i)$.
After performing lengthy derivations, we find that these functions can be
written in the following concise forms 
\begin{eqnarray}
I_{N_D}(\vec{k}'i',\vec{k}i) & = & 
     (\frac{f_{\pi NN}}{m_{\pi}})^2
     \tau_{i'}\gamma^5 \not\! k'
     \frac{1}{2}
     [S_N(p+k) + S_N(p'+k')]
     \tau_i\gamma^5 \not\! k \,,
  \\ 
I_{N_E}(\vec{k}'i',\vec{k}i) & = &    (\frac{f_{\pi NN}}{m_{\pi}})^2
     \tau_{i}\gamma^5 \not\! k
     \frac{1}{2}
     [S_N(p-k') + S_N(p'- k)]
     \tau_{i'}\gamma^5 \not\! k', \label{pinn} \,, \\
 I_{\rho}(\vec{k}'i',\vec{k}i) & = & 
     \frac{ig_{\rho NN}g_{\rho\pi\pi}}{4} \epsilon_{ii'k}\tau_{k}
  \bigg\{ [\gamma_{\mu} - \frac{\kappa_{\rho}}{2m_N}i\sigma_{\mu\nu}
    (p-p')^{\nu} ]
     D_{\rho}^{\mu\lambda}(p-p')(k+k')_{\lambda}] \nonumber \\
   & &     + [(p-p') \leftrightarrow (k'-k)] \bigg\} \,,   \\
I_{\Delta _D}(\vec{k}'i',\vec{k}i) & = &
     (\frac{f_{\pi N\Delta}}{m_{\pi}})^2
     T_{i'}^{\dagger}k_{\mu}'
     \frac{1}{2}
     [ 
       S_{\Delta}^{\mu\nu}(p + k) 
     + S_{\Delta}^{\mu\nu}(p'+ k') \nonumber \\
& &
     - S_{\Delta}^{(+)\mu\nu}(p + k) 
     - S_{\Delta}^{(+)\mu\nu}(p'+ k')
      ] 
     T_{i}k_{\nu} \,, \\
I_{\Delta E}(\vec{k}'i',\vec{k}i) & = &
     (\frac{f_{\pi N\Delta}}{m_{\pi}})^2
     T_{i}^{\dagger}k_{\mu}
     \frac{1}{2}
     [ S_{\Delta}^{\mu\nu}(p - k') 
     + S_{\Delta}^{\mu\nu}(p' - k) ]
     T_{i'}k_{\nu}' \,.
\end{eqnarray}
The propagators in the above equations are defined as 

\begin{eqnarray}
      S_N(p) & = & {1\over \not\! p - m_N}\,, \\
      S_{\Delta}^{\mu\nu}(p) & = &
      {1 \over 3(\not\! p - m_{\Delta})}
     \Bigg [ 2(- g^{\mu\nu}+ \frac{p^{\mu}p^{\nu}}{m_{\Delta}^2})
    + \frac{\gamma^{\mu}\gamma^{\nu} -\gamma^{\nu}\gamma^{\mu}}{2}
    - \frac{p^{\mu}\gamma^{\nu} - p^{\nu}\gamma^{\mu}}{m_{\Delta}} \Bigg ]\,, \\
     D_{\rho}^{\mu\nu}(p)  & = & -
    \frac{g^{\mu\nu} - p^{\mu}p^{\nu}/m^2_{\rho}}{p^2 - m_{\rho}^2}\,. 
\end{eqnarray}
In Eq.\ (3.15), we also have introduced a propagator 
\begin{eqnarray}
S^{(+)\mu\nu}_{\Delta}(p) &=& {m_{\Delta} \over E_{\Delta}(p)} \,
\,{\omega^{\mu}_{p} \bar\omega^{\nu}_{p} \over p_0 - E_{\Delta} (p)},
\end{eqnarray}
where $\omega_{p}^{\mu}$ is the Rarita-Schwinger spinor(as explicitly defined in
Ref.\ \cite{nbl}). In the 
$\Delta$ rest frame, this propagator reduces to the following simple form
\begin{eqnarray}
S^{(+)ij}(p)  \,\,
{\over \vec p \rightarrow 0}\,\,
{1+\gamma^{0}\over 6}\, (3\delta^{ij} - \sigma^{i} \sigma^{j})\,
{1\over p_0 - m_{\Delta}} \,. 
\end{eqnarray}
for $i,j=1,2,3$. The other elements 
involving time components vanish in this special frame; 
$S^{(+)\mu 0} = S^{(+)0\nu} =0$. 
The appearance of this propagator in Eq.\ (3.15) is to remove the
$\pi N \rightarrow \Delta \rightarrow \pi N$ mechanism which can be
generated by the vertex interaction
$\Gamma_{\Delta \leftrightarrow \pi N}$ of the effective Hamiltonian Eq.\ (3.8).
This comes about naturally in our derivations.

We note that the above expressions are remarkbly similar to those
derived from using Feynman rules. The only differences are in the
propagators of the intermediate particles. These propagators are
evaluated by using the momenta of the external particles which 
are restricted on their mass shell, as defined in Eq.\ (3.10). 
For the off-energy-shell dynamics($E_N(k) + E_{\pi}(k) \neq 
E_N(k') + E_{\pi}(k')$), these
propagators can have two possible forms, depending on which set of external
momenta is used. The propagators in Eqs.\ (3.12)-(3.16) are the average of these 
two possible forms of propagators. More details can be seen in Appendix A where
the derivation of $v_N = v_{N_D} + v_{N_E}$ is given explicitly.

In the $\pi N$ center of mass frame, the $\Delta$ in the vertex interaction
$\Gamma_{\Delta\leftrightarrow \pi N}$ is at rest. In this particular frame,
the Rarita-Schwinger spinors reduces to a simple form such that
the matrix element of the vertex interaction 
$\Gamma_{\Delta\leftrightarrow \pi N}$ takes
the following familiar form 
\begin{eqnarray}
<\Delta|\Gamma_{\Delta\leftrightarrow \pi N}|\vec{k}i>
 = -\frac{f_{\pi N\Delta}}{m_{\pi}}\frac{i}{\sqrt{(2\pi)^3}} \,
\frac{1}{\sqrt{2E_{\pi}(k)}} \,
\sqrt{\frac{E_N(k)+m_N}{2E_N(k)}} \,
\vec{S}\cdot \vec{k} T_i.
\end{eqnarray}
Here $\vec{S}$ is a $N\rightarrow\Delta$ transition spin operator. It is
defined by the same reduced matrix element as the transition isospin
operator $T$.

Because of the absence of a $\pi N \leftrightarrow N$ vertex in the
effective Hamiltonian Eq.\ (3.8), it is straightforward to derive
the $\pi N$ scattering equations in the coupled $\pi N \oplus \Delta$ space.
The derivation procedure is similar to that given in Ref.\ \cite{lee86}
for the more complicated $\pi NN$ problem.  The essential idea is to
apply the standard projection operator technique of nuclear reaction
theory \cite{feshbach}.  The resulting scattering amplitude 
can be cast into the following form 
\begin{eqnarray}
T_{\pi N}(E) = t_{\pi N}(E)+\bar{\Gamma}_{\Delta \rightarrow \pi N}(E) 
G_{\Delta}(E)\bar{\Gamma}_{\pi N \rightarrow \Delta}(E).
\end{eqnarray}
The first term is the nonresonant amplitude determined only by the
$\pi N$ potential
\begin{eqnarray}
t_{\pi N}(E) = v_{\pi N} + v_{\pi N} G_{\pi N}(E) t_{\pi N}(E) ,
\end{eqnarray}
with
\begin{eqnarray}
G_{\pi N}(E) =  \frac{P_{\pi N}}{E - E_N(k) - E_\pi (k)+ i \epsilon} , 
\end{eqnarray}
where $P_{\pi N}$ is the projection 
operator for the $\pi N$ subspace.
The second term of Eq.\ (3.23) is the resonant term determined by
the dressed $\Delta$ propagator
and the dressed vertex functions. They are defined by 
\begin{eqnarray}
G_{\Delta}(E)&=& G^0_{\Delta}(E) + G^0_{\Delta}(E)
\Sigma_{\Delta}(E)G_{\Delta}(E) \nonumber \\
             &=&\frac{P_{\Delta}}{E-m_\Delta -\Sigma_{\Delta}(E)}, 
\end{eqnarray}
with
\begin{eqnarray}
G^0_{\Delta}(E)=\frac{P_{\Delta}}{E-m_\Delta} \,, 
\end{eqnarray}
and
\begin{eqnarray}
\bar{\Gamma}_{\pi N \rightarrow \Delta}(E) & = & 
      \Gamma_{\pi N \rightarrow \Delta}( 1+ G_{\pi N}(E)t_{\pi N}(E)), \\
\bar{\Gamma}_{\Delta \rightarrow \pi N}(E) & = & 
      ( 1+ G_{\pi N}(E)t_{\pi N}(E))\Gamma_{\Delta \rightarrow \pi N}, 
\end{eqnarray}
where $P_\Delta$ is the projection operator for the $\Delta$ state, and
the $\Delta$ self-energy is defined by
\begin{eqnarray}
\Sigma_{\Delta}(E) & = & \Gamma_{\pi N\rightarrow\Delta}
G_{\pi N}(E) \bar{\Gamma}_{\Delta\rightarrow\pi N}(E) \,.
\end{eqnarray}

Equations (3.23)-(3.30) are illustrated in Fig.\ 4. These equations are solved 
in partial-wave representation. To find the solution for the integral equation
(3.24), it is necessary to regularize the $\pi N$ potential by introducing
a form factor for each vertex in Eqs.\ (3.12)-(3.16). In this work, we choose
\begin{eqnarray} 
[I_{ND}+I_{NE}](\vec{k}'i',\vec{k}i) &\rightarrow&
[I_{ND}+I_{NE}](\vec{k}'i',\vec{k}i)
F_{\pi NN}(\vec{k}')F_{\pi NN}(\vec{k}) \,, \\
I_{\rho}(\vec{k}'i',\vec{k}i) &\rightarrow& I_{\rho}(\vec{k}'i',\vec{k}i)
F_{\rho NN}(\vec{k}-\vec{k}')F_{\rho \pi\pi}(\vec{k}-\vec{k}') \,, \\
%[I_{\Delta D} + I_{\Delta E} ] (\vec{k}' i', \vec{k} i)
%&\rightarrow & [I_{\Delta D}+I_{\Delta E}](\vec{k}'i',\vec{k}i)
%F_{\pi N \Delta}(\vec{k}')F_{\pi N \Delta}(\vec{k})
\end{eqnarray}
with 
\begin{eqnarray}
F_{\pi NN}(\vec{k})&=&(\frac{\Lambda^2_{\pi NN}}
{\Lambda^2_{\pi NN} + \vec{k}^2})^2 \,, \\
F_{\rho NN}(\vec{k}-\vec{k}') &=& F_{\rho \pi\pi}(\vec{k}-\vec{k}') \,, \\
&=&(\frac{\Lambda^2_{\rho}}
{\Lambda^2_{\rho}+(\vec{k}-\vec{k}')^2})^2 \,.
\end{eqnarray}
For $\pi N \Delta$ vertex with an external pion momentum $\vec{k}$, we 
choose
\begin{eqnarray}
F_{\pi N \Delta}(\vec{k}) =(\frac{\Lambda^2_{\pi N\Delta}}{\Lambda^2_{\pi N\Delta}
+\vec{k}^2})^2.
\end{eqnarray}
%\begin{eqnarray}
%<\Delta | \Gamma_{\Delta \leftrightarrow \pi N}|\vec{k}i> \rightarrow
%<\Delta | \Gamma_{\Delta \leftrightarrow \pi N}|\vec{k}i>
%(\frac{\Lambda^2_{\pi N\Delta}{\Lambda^2_{\pi N\Delta} + \vec{k}^2})^2
%\end{eqnarray}
We have also tried other parameterizations of form factors, but they do not
give better fits to the $\pi N$ scattering  phase shifts.

\section{Pion Photoproduction}

To proceed, we need to first extend the Lagrangian
Eq.\ (3.1) to include $\omega$ meson coupling which is
known \cite{olsson,nbl} to play an important role in pion photoproduction.
We choose the following rather conventional form (with $\kappa_\omega \sim 0$)
\begin{equation}
L_{\omega NN} =
 g_{\omega NN}\bar{\psi}_N(x) 
 [ \gamma_{\mu}  \phi^{\mu}_{\omega}(x)
- \frac{\kappa_{\omega}}{2m_{N}}\sigma_{\mu\nu}\partial^{\nu} 
  \phi_{\omega}^{\mu}(x)] \psi_N(x).
\end{equation}
Following the approach of Ref.\cite{nbl}, the pion photoproduction mechanisms
are defined by the hadronic Lagrangians defined by Eq.\ (3.1) and (4.1) and
the following electromagnetic interaction Lagrangians 
\begin{equation}
L_{em} = L_{\gamma NN} + L_{\gamma\pi\pi} + L_{\gamma\pi N}
    + L_{\gamma\rho\pi} + L_{\gamma\omega\pi}
    + L_{\gamma N \Delta},
\end{equation}
with
\begin{eqnarray}
L_{\gamma NN}& = &
  -e\bar{\psi}_N(x)[
     \hat{e}\vec{A}(x) -
  \frac{\hat{\kappa}}{2m_N}\sigma_{\mu\nu}(\partial^{\nu}A^{\mu}(x))]
\psi_N(x), \\
L_{\gamma\pi NN}& = & 
-\frac{ef_{\pi NN}}{m_{\pi}}
\bar{\psi}_N(x)\gamma_5\vec{A}(x)
(\vec{\tau}\times \vec{\phi}_\pi(x))_3\psi_N(x),    \\
L_{\gamma\pi\pi}& = & 
e(\partial^{\mu}\vec{\phi}_{\pi}(x)\times\vec{\phi}_{\pi}(x))_3A_{\mu}(x), \\
 L_{\rho\pi\gamma}& = & 
 \frac{g_{\rho\pi\gamma}}{m_{\pi}}
 \epsilon_{\alpha\beta\gamma\delta}
 (\partial^{\alpha}A^{\beta}(x))\vec{\phi}_{\pi}(x)\cdot
 (\partial^{\gamma}\vec{\phi}^{\delta}_{\rho}(x)),  \\
 L_{\omega\pi\gamma}& = & 
 \frac{g_{\omega\pi\gamma}}{m_{\pi}}
 \epsilon_{\alpha\beta\gamma\delta}
 (\partial^{\alpha}A^{\beta}(x))\phi^3_{\pi}(x)
 (\partial^{\gamma}\phi^{\delta}_{\omega}(x)),  \\
 L_{\gamma N\Delta} & = &
 ie \bar{\psi}^{\mu}_{\Delta}T_3 \Gamma_{\mu\nu}A^{\nu}(x)
    \psi_N(x) + [h.c.].
\end{eqnarray}
Here $\hat{e} = (1+\tau_3)/2, 
\hat{\kappa}=(\kappa_p+\kappa_n)/2 + (\kappa_p-\kappa_n)\tau_3/2$.  The
$\gamma N\Delta$ coupling in Eq.\ (4.8) is
\begin{equation}
\Gamma_{\mu\nu} = (G_M K_{\mu\nu}^{M} + G_E K_{\mu\nu}^E),
\end{equation}
as defined in Eqs.\ (2.10b,c) of Ref.\ \cite{nbl}.  Its matrix element between an
$N$ with momentum $p$ and a $\Delta$ with momentum $p_{\Delta}$ can be written
explicitly as
\begin{eqnarray}
<\Delta(p_{\Delta})|\Gamma_{\mu\nu}|N(p)>
  =  (G_M - G_E)[ \frac{3}{(m_{\Delta}-m_N)^2 - q^2}
                    \frac{m_{\Delta}+m_N}{2m_N}
                    \epsilon_{\mu\nu\alpha\beta}P^{\alpha}q^{\beta} ]
  \nonumber \\
   +G_E i\gamma^5 [\frac{6}{
            ( (m_{\Delta}+m_N)^2 -q^2)( (m_{\Delta}-m_N)^2 -q^2 )}
           \frac{m_{\Delta}+m_N}{m_N}
 \epsilon_{\mu\lambda\alpha\beta}P^{\alpha}q^{\beta}
 \epsilon^{\lambda}_{ \ \ \nu\gamma\delta}p^{\gamma}_{\Delta}q^{\delta}
\end{eqnarray}
with  $P = (p + p_{\Delta})/2$ and $p_{\Delta}=p+q$.

By appling the usual 
canonical quantization procedure, we can obtain from the above Lagrangians
an electromagnetic interaction Hamiltonian $H_{em}$.
In this work, we will treat the electromagnetic field as an external classical 
field, and hence the electromagnetic interaction $H_{em}$
can be neglected in constructing the unitary transformation operator $S$.
The effective Hamiltonian for describing 
pion photoproduction is therefore a simple extension of the effective Hamiltonian
of the form of Eq.\ (2.21)
\begin{eqnarray}
H_{eff}& \rightarrow& H_{eff} + H_{eff}^{em} \nonumber \\
      &=& H_0 + H^P_I + H^{'P}_I + H^{em}_{eff}
\end{eqnarray}
with
\begin{eqnarray}
H_{eff}^{em} = H_{em} + [H_{em},iS]
\end{eqnarray}
By evaluating $H_{eff}^{em}$ in the coupled $\Delta\oplus \pi N\oplus \gamma N$
subspace, we obtain an extension of Eq.\ (3.8)
\begin{eqnarray}
H_{eff}^{\pi} \rightarrow H_{eff}^{\gamma\pi}= H_0+\Gamma_{\Delta\leftrightarrow\pi N}
+v_{\pi N} + \Gamma_{\Delta\leftrightarrow\gamma N} + v_{\pi\gamma},
\end{eqnarray}
where $\Gamma_{\Delta \leftrightarrow \pi N}$ and $v_{\pi N}$ have already been
given in Eqs.\ (3.11)-(3.22). The resonant and nonresonant electromagnetic
interactions are, respectively, described by 
$\Gamma_{\Delta \leftrightarrow \gamma N}$ and $v_{\pi\gamma}$, and are illustrated
in Fig.\ 5. We again omit the details of the derivation of these two terms, and
simply present our results in the center of mass frame. The momenta variables
$q^{\mu}$ for the photon, $p^{\mu}$ for the initial nucleon,
$k^{\mu}$ for the pion, and $p^{'\mu}$ for the final nucleon in Fig.\ 5 are therefore

\begin{eqnarray}
q^\mu &=& ( q,\vec{q}), \nonumber \\
p^\mu &=& (E_N(q),-\vec{q}), \nonumber \\
k^\mu &=& (E_\pi(k), \vec{k}), \nonumber \\
p^{'\mu} &=& (E_N(k),-\vec{k}).
\end{eqnarray}
In terms of these variables, the expression Eq.\ (4.10)
becomes very simple. The matrix element of the
$\Gamma_{\Delta\leftrightarrow \gamma N}$ vertex can then be expressed in terms of
the spin and isospin $N\rightarrow \Delta$ transition operators $S$ and $T$
introduced in section III.  At the resonance energy, $m_{\Delta} = E_N (q) + q$,
Eq.\ (4.10) leads to
\begin{eqnarray}
<\Delta|\Gamma_{\gamma N\rightarrow\Delta}|\vec{q}\lambda>
 &=& - \frac{e}{(2\pi)^{3/2}} \sqrt{\frac{E_N(q)+m_N}{2E_N(q)}}
\frac{1}{\sqrt{2q}} \frac{3m_{\Delta}}{2m_N(m_{\Delta}+m_N )}
  \nonumber \\
& &
T_3 [ i G_M \vec{S}\times \vec{q} \cdot \vec{\epsilon}_\lambda
      + G_E ( \vec{S}\cdot\vec{\epsilon}_\lambda \vec{\sigma}\cdot\vec{q}
            + \vec{S}\cdot \vec{q} \vec{\sigma}\cdot\vec{\epsilon}_\lambda) ],
\end{eqnarray}
where $\vec{\epsilon}_{\lambda}$ is the photon polarization vector.
The matrix element of the nonresonant interaction $v_{\pi\gamma}$
can be written as
\begin{eqnarray}
<\vec{k}i,m_s'm_\tau'|v_{\pi\gamma}|\vec{q}\lambda,m_s m_\tau>
&=&\frac{1}{(2\pi)^3}\frac{1}{\sqrt{2E_{\pi}(k)}}\frac{1}{\sqrt{2q}}
\sqrt{\frac{m_N}{E_N(k)}}\sqrt{\frac{m_N}{ E_N(q)}} \nonumber \\
&\times& \bar{u}_{-\vec{k}m_s'm_\tau'}
     [\sum_{\alpha}I^{\pi\gamma}_{\alpha}(\vec{k}i,\vec{q}\lambda)]
   u_{-\vec{k}m_sm_\tau}.
\end{eqnarray}
The nonresonant pion photoproduction mechanisms
are contained in  $I_N^{\pi\gamma}$ for the direct nucleon terms (Figs.\ 5c,5d,5e) 
, $I_{\pi}^{\pi\gamma}$ for the pion pole 
term (Fig.\ 5f), $I_{\rho,\omega}^{\pi\gamma}$
for the vector meson exchange (Fig.\ 5g), and 
$I^{\pi\gamma}_{\Delta D,\Delta E}$
for the direct and exchange $\Delta$ terms (Figs.\ 5h,5i). Explicitly, we have
\begin{eqnarray}
I^{\pi\gamma}_{N}(\vec{k}i,\vec{q}\lambda) &= &
\frac{ef_{\pi NN}}{m_{\pi}}  [ 
  i\tau_i\gamma^5\not\! k S_N(p'+k)
 ( \hat{e}\not\! \epsilon (\lambda) + i\frac{\hat{\kappa}}{2m_N}
   \sigma_{\mu\nu}\epsilon(\lambda)^{\mu}q^{\nu} )  \nonumber \\
 & &+ i(\hat{e}\not\! \epsilon(\lambda) + i\frac{\hat{\kappa}}{2m_N}
   \sigma_{\mu\nu}\epsilon(\lambda)^{\mu}q^{\nu} )S_N(p-k)
\tau_i\gamma^5\not\! k
  - \epsilon_{ij3}\tau_j\gamma^5\not\! \epsilon (\lambda) ] 
   \\
I^{\pi\gamma}_{\pi}(\vec{k}i,\vec{q}\lambda) &= & 
 - \frac{ef_{\pi NN}}{m_{\pi}}
 \epsilon_{ij3}\tau_j\gamma^5 (\not\! p' - \not\! p)
 ((k+p-p') \cdot \epsilon(\lambda)D_{\pi}(p-p')
    \\
I^{\pi\gamma}_{\rho}(\vec{k}i,\vec{q}\lambda) &= &
 \frac{g_{\rho NN}g_{\rho\pi\gamma}}{m_{\pi}}
  \frac{\tau_i}{2}
  [ \gamma_{\mu} -
  \frac{i\kappa_{\rho}}{2m_N}\sigma_{\mu\eta}(p-p')^{\eta} ] \nonumber 
\times D^{\mu\nu}_{\rho}(p-p')\epsilon_{\alpha\beta\gamma\nu}
   \epsilon^{\alpha}(\lambda) q^{\beta} (p-p')^{\gamma}
    \\
I^{\pi\gamma}_{\omega}(\vec{k}i,\vec{q}\lambda) & = &
 \frac{g_{\omega NN}g_{\omega\pi\gamma}}{m_{\pi}}
  \tau_3
  [ \gamma_{\mu} -
  \frac{i\kappa_{\omega}}{2m_N}\sigma_{\mu\eta}(p-p')^{\eta} ] 
 \times D^{\mu\nu}_{\omega}(p-p')\epsilon_{\alpha\beta\gamma\nu}
   \epsilon^{\alpha}(\lambda) q^{\beta} (p-p')^{\gamma}
    \\
I^{\pi\gamma}_{\Delta E}(\vec{k}i,\vec{q}\lambda)& = &
 \frac{ef_{\pi N\Delta}}{m_{\pi}}T^{\dagger}_3 
 \Gamma^{\dagger}_{\mu\delta}\epsilon^{\delta}(\lambda) S_{\Delta}^{\mu\nu}(p-k)T_i k_{\nu} 
    \\
I^{\pi\gamma}_{\Delta D}(\vec{k}i,\vec{q}\lambda)& = &
- \frac{ef_{\pi N\Delta}}{m_{\pi}}T^{\dagger}_i k_{\mu} 
 [ S_{\Delta}^{\mu\nu}(p'+k) - S^{(+)\mu\nu}(p'+k) ]
 T_3 \Gamma_{\nu\delta}\epsilon^{\delta}(\lambda)
\end{eqnarray}
Here we observe again that the above expressions are 
very similar to the results derived by using Feynman rules.
However, they have an important feature that 
the time components of the momenta in the
propagators and strong interaction vetices are evaluated by 
using the external momenta of the final $\pi N$ state. This is the consequency of
appling the unitary transformation method defined in Eq.\ (4.12).
In addition to including nonresonant $\Delta$ terms (Figs.\ 5h and 5i), this is
an another feature which makes our model different from the model developed
in Ref.\ \cite{nbl}.

It is straightforward to derive from the effective Hamiltonian Eq.\ (4.13)
the t-matrix of pion photoproduction
\begin{eqnarray}
T_{\gamma\pi}(E)  =  t_{\gamma\pi}(E) +
 \bar{\Gamma}_{\Delta\rightarrow\pi N}(E)G_{\Delta}(E)
 \bar{\Gamma}_{\gamma N\rightarrow\Delta}(E)\,,
\end{eqnarray}
where the nonresonant amplitude is defined by
\begin{eqnarray}
t_{\gamma\pi}(E)= v_{\gamma \pi} +  t_{\pi N}(E)G_{\pi N}(E) v_{\gamma\pi}\,.
\end{eqnarray}
The dressed $\gamma N\leftrightarrow \Delta $ is defined by
\begin{eqnarray}
\bar{\Gamma}_{\gamma N\rightarrow\Delta}(E)
=\Gamma_{\gamma N \rightarrow \Delta} 
+ \bar{\Gamma}_{\pi N\rightarrow\Delta}(E) G_{\pi N}(E)v_{\gamma \pi} \,.
\end{eqnarray}
In the above equations, $G_{\Delta}, 
\bar{\Gamma}_{\Delta \leftrightarrow \pi N}, G_{\pi N}$ and $t_{\pi N}$ have
been defined in section III. The standard partial-wave decomposition is used
to obtain the multipole amplitudes from $T_{\gamma\pi}$ for the 
$\gamma N \rightarrow \pi N$ reaction, and from $\bar{\Gamma}_{\Delta 
\leftrightarrow \pi N}$ for the dressed $\Delta \leftrightarrow \pi N$ vertex.
Eqs.(4.22)-(4.24) are illustrated in Fig.6.

The $K$ matrix formulation of the 
$\gamma N \rightarrow \pi N$ reaction is often used
\cite{olsson,davidson91,davidson90,li}
in the analysis of data.  Within our formulation, this can be obtained 
by replacing the $\pi N$ free Green function $G_{\pi N}$
Eq.\ (3.25) by 
\begin{eqnarray}
G_{\pi N}(E) \rightarrow G^{P}_{\pi N}(E) 
= P \frac{P_{\pi N}}{E-E_N(k)-E_{\pi}(k)},
\end{eqnarray}
where $P$ means taking the principal-value part of the propagator. If
this replacement is used in the calculations of Eqs.\ (3.23)-(3.30), 
all scattering 
quantities will be real numbers. These $K$-matrix quantities are defined by
exactly the same Eqs.\ (3.23)-(3.30) with the following changes
\begin{eqnarray}
G_{\pi N} &\rightarrow& G^P_{\pi N}\,, \nonumber \\
T_{\pi N} &\rightarrow& K_{\pi N}\,, \nonumber \\
t_{\pi N} &\rightarrow& k_{\pi N}\,, \nonumber \\
\bar{\Gamma}_{\Delta\leftrightarrow\pi N} &\rightarrow& 
\bar{\Gamma}^k_{\Delta\leftrightarrow\pi N} \,, \nonumber \\
G_{\Delta}&\rightarrow& G^P_{\Delta} = \frac{P_{\Delta}}{E-m_\Delta -
\Sigma^k_{\Delta}(E)}  \,,
\end{eqnarray}
with
\begin{eqnarray}
\Sigma^k_{\Delta}(E)= \Gamma_{\pi N\rightarrow\Delta}G^P_{\pi N}(E) 
\bar{\Gamma}^k_{\Delta\rightarrow\pi N}(E) \,.
\end{eqnarray}
Note that the $\Delta$ self-energy $\Sigma^k_{\Delta}$ is now a real number,
and the propagator $G^P_{\Delta}$ has a pole at $E = M_R = m_\Delta +
\Sigma^k_\Delta(M_R)$.

The corresponding K-marix for pion
photoproduction can be obtained from Eqs.\ (4.22)-(4.24) by the same replacement
Eq.\ (4.25)
\begin{eqnarray}
K_{\gamma \pi}(E) = k_{\gamma\pi}(E) +
\bar{\Gamma}^k_{\Delta\rightarrow\pi N}(E) G^P_{\Delta}(E)
\bar{\Gamma}^k_{\gamma N\rightarrow\Delta}(E) \,,
\end{eqnarray}
with 
\begin{eqnarray}
k_{\gamma\pi} &=& v_{\gamma\pi} +  k_{\pi N}(E)G^P_{\pi N}(E)v_{\gamma\pi} \,, \\
\bar{\Gamma}^k_{\gamma N\rightarrow\Delta}(E)&=& \Gamma_{\gamma N\rightarrow\Delta}
+  \bar{\Gamma}^k_{\pi N\rightarrow\Delta}(E)G^P_{\pi N}(E)v_{\gamma \pi} \,.
\end{eqnarray}
For the on-shell matrix elements($E=E_N(k_0)+E_\pi(k_0)=q + E_N(q)$), 
it is straigtforward to find the following relation in each partial wave
\begin{eqnarray}
T_{\gamma \pi}(k_0,q) = ( 1-i\pi\rho T_{\pi N}(k_0,k_0))
K_{\gamma \pi}(k_0,q) \,.
\end{eqnarray}

For investigating the hadron structure, we are interested in the $\gamma N
\leftrightarrow \Delta$ vertex. As seen in Eqs.\ (4.24) and (4.30), the dressed
vertices in t-matrix and in K-matrix are different. In the t-matrix formulation
$\bar{\Gamma}_{\gamma N,\Delta}$ Eq.\ (4.24) is a complex quantity, while in 
the K-matrix formulation $\bar{\Gamma}^k_{\gamma N,\Delta}$, Eq.\ (4.30), is a
real function. Consequently, we need to be careful about the meaning of
the $E2/M1$ ratio of the dressed $\gamma N \leftrightarrow \Delta$ vertex. The
clearest definition seems to be in the K-matrix formulation because as 
the energy
approaches the resonance position $M_R=m_\Delta +\Sigma^k_{\Delta}(M_R)$, 
Eq.\ (4.28) is reduced to
\begin{eqnarray}
K_{\gamma \pi}(E) = \frac{A}{E- M_R} + B
\end{eqnarray}
with 
\begin{eqnarray}
A &=& \bar{\Gamma}^{k}_{\Delta\rightarrow\pi N}(E)
      \bar{\Gamma}^{k}_{\gamma N\rightarrow\Delta}(E)
\\
B &=& k_{\gamma\pi}(E)
\end{eqnarray}
The separable form of the residue $A$ of the K-matrix leads to an interesting
result that the ratio between the $E1$ and $M1$ multipole amplitudes of the 
dressed $\gamma N \Delta$ vertex can be directly calculated from the residues of
the corresponding multipole amplitudes of the $\gamma N \rightarrow \pi N$
reaction. The reason is that both amplitudes have the same strong interaction
dressed vertex in the $P_{33}$ channel, and hence the ratio between the residues
does not depend on it. Explecitly, we have
\begin{eqnarray}
R_{EM}&=& [\frac{A(E_{1^+})}{A(M_{1^+})}]_{\gamma N \rightarrow \pi N} \nonumber
\\
 &=& \frac{\bar{\Gamma}^k_{\Delta\rightarrow\pi N}(P_{33})
\bar{\Gamma}^{k}_{\gamma N\rightarrow\Delta}(E_{1^+})}
{
\bar{\Gamma}^k_{\Delta\rightarrow\pi N}(P_{33})
\bar{\Gamma}^{k}_{\gamma N\rightarrow\Delta}(M_{1^+})} \nonumber \\
&=&\frac{\bar{\Gamma}_{\gamma N\rightarrow\Delta}(E_{1^+})} 
{\bar{\Gamma}_{\gamma N\rightarrow\Delta}(M_{1^+})}
\end{eqnarray}
The above relation is the basis of the model-independent
analysis of Ref.\ \cite{davidson90}.  We will discuss this issue in the next section.

\section{Results and Discussions}

Our first task is to determine the parameters of
the effective $\pi N$ Hamiltonian derived in section II.
Apart from the known $\pi NN$ coupling
constant, $\frac{f^2_{\pi NN}}{4\pi}=0.08$, the model has seven parameters:
the couplig constants ($g_{\rho}^2=g_{\rho NN}g_{\rho\pi\pi}$, $\kappa_{\rho}$, 
$f_{\pi N\Delta}$) 
of Eqs.\ (3.12)-(3.16), the cutoff parameters ($\Lambda_{\pi NN},
\Lambda_{\pi N\Delta}, \Lambda_{\rho}$) of the form factors Eqs.\ (3.34)-(3.37),
and $m_{\Delta}$ of the $\Delta$ bare mass.
These parameters are determined by fitting the $\pi N$
phase shifts. Without including inelastic channels, our scattering equations
Eqs.\ (3.23)-(3.30) are valid rigorously only in the energy region 
where the $\pi N$ scattering is purely elastic.
We therefore first take a conservative approach to only fit the data in the 
energy region below $T_L = 250$ MeV pion laboratory energy. This model, called Model-L,
is sufficient for investigating 
pion photoproduction up to 400 MeV photon laboratory energy.
Our results are displayed in Fig.\ 7. 
We see that within the uncertainties of the phase shift data \cite{arndt,hohler,koch}
the model can give a good account of all s- and p- partial waves except the
$P_{13}$ channel at $T_L > 120$ MeV. We have found that this difficulty can not be removed by
tring various form factors other than those given
in Eqs.\ (3.34)-(3.37), and following the previous works\cite{pearce,yang1}
to include the exchange of a fictituous scalar 
$\sigma$ meson. To see the origin of this problem,
we show  in Fig.\ 8 the
contributions from each mechanism of Fig.\ 3 
to the on-shell matrix elements of the $\pi N$
potential. Clearly, the fit to the phase shift data involves delicate
cancellations between different mechanisms. It is possible to
imporve the fit to $P_{13}$ by weakening the $\rho$-exchange or
the $\Delta$-exchange. But this change will destroy the good fits to all other
partial waves. Fortunately, the $\pi N$ scattering effect due to the
$P_{13}$ channel is weak
in determining the pion photoproduction cross sections. We therefore will
not pursue the solution of this problem here. Perhaps this can be solved only
when the $\rho$-exchange is replaced by the two-pion-exchange considered 
in Ref.\ \cite{julich2}. To be consistent, the coupling with two-pion channels, 
such as $\pi \Delta$ and $\rho N$, must also be included.  These two
possible improvements can be achieved by
extending the unitary transformation method introduced in Sections II-IV to second-order
in the coupling constants.  

Let us now examine in more detail the $P_{33}$ channel which is most relevant to
our later investigation of the $\Delta$ excitation in pion photoproduction.  
As seen in Eq.\ (3.23), the resonant part of the
$P_{33}$ amplitude is determined by the dressed propagator $G_{\Delta}$ of
Eq.\ (3.26) and the dressed vertex $\bar{\Gamma}_{\Delta \leftrightarrow \pi N}$ 
of Eqs.\ (3.28) and (3.29). Clearly, 
the $\Delta$ resonance peak of $\pi N$ scattering can be obtained only 
when the model can generate a $\Delta$ self-energy such that 
the real part of $(E-m_\Delta - \Sigma_\Delta(E)) \rightarrow 0$ as
the $\pi N$ invariant mass $W$ approaches the resonance enery 
$W = M_R =1236$ MeV. Our model has this
desired property, as illustrated in the upper half of Fig.\ 9.
Another important feature in the $P_{33}$ channel
is that the $\pi N$ potential generates the dressed 
$\Delta \leftrightarrow \pi N$ vertex, as defined in Eqs.\ (3.28) and (3.29).
We have found that this renormalization effect modifies greatly 
the $\Delta \leftrightarrow \pi N$ form factor in the low momentum region.
To see this, we cast the 
bare vertex (Eq.\ (3.22) including the form factor $F_{\pi N\Delta}(k)$ 
defined by Eq.\ (3.37)) and the dressed vertex (Eq.\ (3.28)) 
into the following forms 
\begin{eqnarray}
\langle\Delta|\Gamma_{\Delta\leftrightarrow \pi N}|\vec{k}i\rangle
  = -\frac{f_{\pi N\Delta}}{m_{\pi}}\frac{i}{\sqrt{(2\pi)^3}}\,
    \frac{1}{\sqrt{2m_{\pi}}}
    F_{bare}(k)\vec{S}\cdot \vec{k} T_i.
\end{eqnarray}
\begin{eqnarray}
\langle\Delta|\bar{\Gamma}_{\Delta\leftrightarrow \pi N}|\vec{k}i\rangle
  = -\frac{\bar{f}_{\pi N\Delta}}{m_{\pi}}\frac{i}{\sqrt{(2\pi)^3}}\,
    \frac{1}{\sqrt{2m_{\pi}}}
    F_{dressed}(k)\vec{S}\cdot \vec{k} T_i.
\end{eqnarray}
with the normalization $|F_{dressed}(0)| = F_{bare}(0)=1$.
We find that the dressed coupling constant, 
$\bar{f}_{\pi N \Delta}$, is 1.3 of the bare
coupling constant $f_{\pi N \Delta}$. The dressed form factor 
$\bar{F}_{\pi N\Delta}(k)$ 
falls off faster than the bare form factor $F_{\pi N\Delta} (k)$
in momentum space, as seen in the 
lower half of Fig.\ 9. This means that 
the nonresonant $\pi N$ interaction has extended 
the $\Delta$ excitation region to a larger distance in coordinate space.
 
A significant difference between our approach and the previous $\pi N$ 
models\cite{pearce,yang1,julich2,surya1}
is in the treatment of $P_{11}$ channel. By employing the unitary transformation,
the $\pi N \leftrightarrow N$ vertex does not appear in the 
effective Hamiltonian Eq.\ (3.8) and hence our formulation of $\pi N$ 
scattering is straightforward. It does not require the nucleon mass 
renormalization. It is natual to ask whether our approach
possesses the well-established nucleon-pole dynamics. 
This question can be answered by
examining Fig.\ 10 in which the $\pi N$ phase shifts and the scattering t-matrix
elements calculated from the nucleon pole term $v_{N_D}$ only (Fig.\ 3c)  and
the full potential are compared. We see in the the upper half of Fig.\ 10
that $\pi N$ phase shifts due to the
nucleon-pole term (dotted curve) are repulsive as expected.
The fit to the $P_{11}$ data is due to a delicate cancellation between 
the repulsive nucleon-pole term and the attraction coming mainly from 
$\rho-$ and $\Delta-$exchange terms(see the $P_{11}$ case in Fig.\ 8). 
In the lower half of Fig.\ 10, we see that as the energy
approaches the threshold, $W=m_\pi+m_N$, the nucleon pole term(dotted curve) 
apparently dominates the interaction.
If we analytically continue to the nucleon pole position,  
$k_0=i k_N$($ m_N=E_N(ik_N)+E_\pi(ik_N)$), the
scattering amplitude will be determined by the nucleon pole term, i.e. 
$t(k_0,k_0, W\rightarrow m_N)\sim  v_{N_D} \sim 1/(W-m_N)$.
 
The parameters of the constructed model 
are listed in the first row (Model-L) of Table I. 
The calculated scattering lengths are presented in the first column of Table II. 
They are all in good agreement with the data.
If we assume the universality of $\rho$ coupling, we then
have $g_{\rho NN} = g_{\rho\pi\pi} = 6.2$ which is close to
that determined in Refs.\ \cite{pearce,yang1}. The fit is also sensitive to 
the $\rho$ tensor coupling constant $\kappa_\rho$. Our value is close to that of
Ref.\ \cite{pearce}, but is much smaller than 6.6 used in Ref.\ \cite{yang1}. 

We now turn to presenting our results of pion photoproduction. 
With the $\pi N N, \pi\Delta N$ and $\rho NN$ vertices 
defined by the parameters given in Table 1, 
the considered pion photoproduction mechanisms (Fig.\ 5) still have 
unknown parameters associated with the vector meson-exchange and
the $\gamma N \leftrightarrow \Delta$ vertex. Following the 
previous approach \cite{nbl}, we assume
that the photon-meson coupling constants $g_{\rho\pi\gamma}$
and $g_{\omega\pi\gamma}$ can be determined from the partial decay widths listed
by the Particle Data Group \cite{particle}. For the $\omega$ meson, we further
assume that the tensor coupling 
$\kappa_{\omega NN}=0$ and the $\omega NN$ form factor is identical to the
$\rho NN$ form factor given in Table 1. The coupling constant $g_{\omega NN}$ is
not well determined in the literature. We will treat it as a free parameter,
although the quark model value $g_{\omega NN} = (3g_{\rho NN})/2$ seems to
be a reasonalbe guess. Thus, 
our investigation of pion photoproduction has only three adjustable parameters:
$G_M$ and $G_E$ of the bare $\Delta \leftrightarrow \gamma N$ 
vertex, and the coupling
constant $g_{\omega NN}$ of the $\omega$ exchange. 
We have, however, some ideas about the ranges of these parameters.
If we assume that bare vertex interaction $\Gamma_{\Delta \leftrightarrow\gamma N}$ can be 
identified with the 
constituent quark model \cite{isgur,bijker,capstick,close,keister}, 
then $|G_E/G_M| \sim 0$ since the one-gluon-exchange
interaction gives negligible $D$-state components 
in $N$ and $\Delta$. We also expect that the $\omega$
coupling should be close to the quark model prediction, 
$g_{\omega NN} = 3g_{\rho NN}/2 \sim 9$, if the $\rho$ coupling from our 
$\pi N$ model (Table I) is used. It is therefore reasonable to only consider
the region  $g_{\omega NN} \le$ 15 and $|G_E/G_M |\le 0.1$.

Since the $\omega$-exchange mechanism (Fig.\ 5g) does not produce charged pions
directly (only through $\pi N$ charge exchange), the 
ranges of $g_{\omega NN}, G_M$ and $G_E$
can be most sensitively determined by considering the 
data of $\pi^0$ photoproduction. 
In the considered region 
that $|G_E/G_M| \le 0.1$ and $g_{\omega NN} \le $ 15, we
have found that the magnitudes of the $\pi^0$ differential cross 
sections depend mainly on $G_M$. The values of $g_{\omega NN}$ and
$G_E$ can be narrowed down by considering spin observables. In this work,
we make use of the recent LEGS \cite{legs} data
of photon-asymmetry ratios $R_\gamma =d\sigma_{\parallel}/d\sigma_{\perp}$
of $\gamma p \rightarrow \pi^0 p$ reaction.
We have found that the slope of $R_\gamma(E)$ at a fixed pion angle 
is sensitive to the value of $g_{\omega NN}$. This is illustrated in the
upper half of Fig.\ 11 for the case of $G_M=1.85$ and $ G_E=0.025$. A smaller
$g_{\omega NN}$ yields a steeper slope. The 
data clearly favor $g_{\omega NN} \sim 10.5$. 
In the lower half of Fig.\ 11, we see that the magnitude, not the slope, of
$R_\gamma$ is significantly 
changed by varing the value of $G_E$ from -0.1 to +0.1.
The data are consistent with 
$-0.025 \le G_E \le 0.025$, while $G_E = +0.025$ seems to give a better
fit. Results similar to Fig.\ 11 can be obtained
by using a higher value $G_M=1.95 $. In this case, a smaller value 
of $g_{\omega NN} = 7$ is needed to maintain the same fit to the magnitude
of the differential cross section as well as the slope of the 
$R_\gamma$. But the best value of $G_E$ to reproduce the 
magnitude of $R_\gamma$ is $-0.025$
instead of $+0.025$ for the $g_{\omega NN}=10.5$ case.
In fact, we have observed 
a strong correlation between the allowed values
of $G_M$ and $g_{\omega NN}$. In all cases, the allowed value of $G_E$ is consistent
with $-0.025 \le G_E \le +0.025$. Therefore, the acceptable
values of ($G_M,g_{\omega NN}$) are
on the curve between the $G_E=-0.025$ and $G_E=+0.025$ lines in Fig.\ 12.
To determine the precise value of $R_{EM}=\frac{G_E}{G_M}$, 
which measures the deformation of the $\Delta$,
we clearly need to pin down the $\omega$ meson coupling constant $g_{\omega NN}$.

The predictions from using the parameters lying on the curve 
between $G_E = + 0.025$ and $G_E = - 0.025$ lines of Fig.\ 12 are also
in good agreement with the $\pi^0$ data at other angles and the $\pi^+$ data.
These are shown in Fig.\ 13 for $\pi^0$ production and Fig.\ 14 for $\pi^+$
production. These results are obtained from using the parameters defined by
the interaction points of the $G_E=\pm 0.025$ lines in Fig.\ 12: 
$(g_{\omega NN}, G_M, G_E) = (10.5, 1.85,+0.025)$, and $(7., 1.95, -0.025)$.
Both sets of parameters yield equally good agreements with 
the $\pi^0$ data (Fig.\ 13). For $\pi^+$ production,
the predictions are in good agreements with the data of 
the photon-asymmetry ratios
$R_\gamma$, but underestimate the differential cross sections by 
about 10 percent at
most energies. Since the $\omega$ exchange has a small contribution to
the $\pi^+$ production (only through
charge-exchange $\pi N$ final state interaction), the only way 
to resolve this difficulty within our model is to increase the value of $G_M$. 
But this will lead to an overestimate of the $\pi^0$ cross section from Bonn.

Although the difficulty in reproducing the $\pi^+$ data in Fig.\ 14b 
could be an indication of the deficiency of our model, the
possibility of a larger $\pi^0$ cross section has been suggested by 
three $\pi^0$ data at $\theta=120$ from Ref.\ \cite{dugan} (Fig.\ 13b).
To fit these three data points, we need to 
increase $G_M$ from 1.85 to 2.0 for the case of $g_{\omega NN}$ = 10.5 and 
from 1.95 to 2.1  for the case of $g_{\omega NN}$ = 7. The results from these two
changes in $G_M$ are, respectively, the solid
and dotted curves in Figs.\ 15 and 16. The predicted
photon-asymmetry ratios (Figs.\ 15a and 16a) are still in good agreement with the 
data. The agreement with the $\pi^+$ data (Fig.\ 16b) is clearly improved.
But the calculated $\pi^0$ differential cross sections (Fig.\ 15b) 
overestimate the Bonn data \cite{sandorfi,menze} by about 15 percent. 
Clearly, the disagreement between the 
$\pi^0$ data at $\theta=120^0$ (Fig.\ 15b) from Refs.\ \cite{sandorfi,menze} and \cite{dugan} 
must be resolved by new measurements.

To further reveal the dynamical content of our model, we compare 
in Fig.\ 17 our predictions of angular distributions with the data compiled in
Ref.\ \cite{menze} for $\gamma p \rightarrow \pi^0 p$ (Fig.\ 17a),
$\gamma p \rightarrow \pi^+ n$ (Fig.\ 17b), and $\gamma n \rightarrow \pi^- p$
(Fig.\ 17c) reactions. 
All results calculated from using the parameters lying on 
the curve between the $G_E = + 0.025$ and $G_E = - 0.025$ lines in Fig.\ 12 are 
very close to the solid curves which are from using 
$(g_{\omega NN}, G_M, G_E)=(10.5,1.85,0.025)$.  Again, we see that the
charged pion production cross sections are underestimated. If a larger $G_M=2.0$
is used in this calculation, we obtain the dotted curves which are in
a better agreement with the charged pion data, but overestimate the $\pi^0$ data
by about 15 percent at resonance peaks. In all cases, the theoretical
predictions underestimate the data at 380 MeV and higher energies. 
This is expected, since
the constructed model does not include inelastic channels which should start
to play a significant role at energies above about 350 MeV. For example, 
the inelastic production mechanism 
$\gamma N \rightarrow \pi \Delta \rightarrow \pi N$
should exist since it is known that the $\pi N$ scattering at this
higher energy can be described only when the coupling with the 
$\pi \Delta$ channel is included.
To investigate this effect, it is necessary to
extend the derivation of effective Hamiltonians presented
in sections III and IV to include the $\pi\Delta$ as well as other 
two-pion states. 

We now focus on the theoretical interpretations of the
$\Delta \leftrightarrow \gamma N$ vertex. The values of $G_M$ and $G_E$ 
determined above characterize the bare $\Delta \leftrightarrow \gamma N$ vertex 
which can only be identified with hadron models with no coupling with the 
$\pi N$ or other hadronic reaction channels. One possible interpretation is
to compare the determined $G_M$ and $G_E$ with the predictions of 
the most well-developed constituent quark model \cite{isgur,bijker,capstick,close,keister}. 
To explore this possibility, it is necessary to first
discuss the quantities in our model which can be compared with
the results from empirical amplitude analyses\cite{davidson90,li,berends}.
For investigating
the $\Delta$ mechanism, we need to only consider 
the $\gamma N \rightarrow \pi N$ multipole amplitudes $M_{1^+}$ 
and $E_{1^+}$ with a $P_{33}$ final $\pi N$ state, 
and the dressed vertex function $\bar{\Gamma}_{\gamma N,\Delta}$. 
These can be computed
from Eqs.\ (4.22)-(4.24) or Eqs.\ (4.28)-(4.30) by performing the standard partial-wave
decomposition (see, for example, the appendix of Ref.\ \cite{nbl}). 
We will discuss these quantities using
the results calculated from setting 
$(g_{\omega NN}, G_M, G_E)=(10.5, 1.85, 0.025)$ (solid
curves in Figs.\ 13, 14 and 17).

The predicted amplitudes $M_{1^+}$ and $E_{1^+}$ are compared in Fig.\ 18 with the results
from the empirical amplitude analyses Ref.\ \cite{li,berends}. We see 
in the upper part of Fig.\ 18 that
the predicted $M_{1^+}$ amplitudes are in good agreement with empirical values.
In the lower half, we show that both the $E_{1^+}$ amplitudes calculated from using
$G_E=+0.025$ (solid curves) and $G_E=-0.025$ (dotted curves) are within the
uncertainties of the amplitude analyses.  This is consistent with our analysis
using LEGS data, as seen in the lower part of Fig.\ 11. 
The uncertainties of the empirical values of the $E_{1^+}$  amplitude
are due to the lack of complete data of spin observables.
More experimental efforts are clearly needed to pin down the value of $G_E$ which
is needed to test models of hadron structure.

The dressed $\Delta \leftrightarrow \gamma N$ vertex, defined by Eq.\ (4.24), is
a complex number. By making the usual partial-wave decomposition, its
magnetic M1 and electric E2 components can be written as
$\Gamma(\alpha) = |\Gamma(\alpha)| e^{i\phi(\alpha)}$ with 
$\alpha = M_{1^+}, E_{1^+}$.
The predicted dressed vertex functions 
$\Gamma(\alpha)$ are the solid curves in Fig.\ 19.
We see that their magnitudes $|\Gamma(\alpha)|$ are very different from 
the corresponding values(dotted curves) 
of the bare $\Delta \leftrightarrow \gamma N$ vertex.
The differences are due to the very large contribution of the
nonresonant mechanism described by the second term of Eq.\ (4.24).
Our results indicate that an accurate reaction theory
calculation of the nonresonant pion photoproduction
mechanisms is needed to determine the bare $\Delta \leftrightarrow \gamma N$
vertex from the pion photoproduction data. This requires a dynamical treatment of
the nonresonant pion photoproduction mechanisms,
as we have done in this work. 
Within the meson-exchange formulation presented in this work, the determined 
$G_M$ and $G_E$ of the bare $\Delta \leftrightarrow \gamma N$ vertex can be compared with
the predictions from a hadron model which does not include the coupling with
the $\pi N$ ``reaction" channel (both pion and nucleon are on their mass-shell). 

We now turn to investigate the $K$-matrix method which has been the 
basis of the empirical amplitude analyses of Refs.\ \cite{davidson90,li}.
In Ref.\ \cite{davidson90}, it was shown that if the 
background term is assumed to be a slowly varing function of energy,
the ratio $R_{EM}$ between the $E_{1^+}$ and $M_{1^+}$ of
the $\gamma N \rightarrow \Delta$ transition       
at the resonant energy $W=M_R$ 
can then be extracted ``model-independently"
from the data of the $\gamma N\rightarrow \pi N$ reaction.
The only limitation is the accuracy of the employed $\pi N$ amplitudes
and the $M_{1^+}$ and $E_{1^+}$ multipole amplitudes of the  
$\gamma N\rightarrow \pi N$ reaction.  By using Eqs.\ (4.32)-(4.35), we can
examine whether the $K$-matrix method of Ref.\ \cite{davidson90} is consistent
with our dynamical model.  In Fig.\ 20, we display
our predictions of the energy-dependence of the total K-matrix (solid curve),
the contribution from the resonant term (dotted 
curve) which has a pole at $W=1236$ MeV, and the
nonresonant contribution B (dashed curve). Clearly, the 
energy-dependence of the nonresonant term $B$ is rather weak.  The
assumption made in the empirical 
analysis of Ref.\ \cite{davidson90} is fairly consistent with our dynamical model.

By using Eq.\ (4.33), we can calculate the residue $A$ of the $K$-matrix
from the dressed vertex 
$\bar{\Gamma}^k_{\gamma N\rightarrow\Delta}$ defined by Eq.\ (4.30).
The results (solid curve) for the $M_{1^+}$ transitions are 
compared with that calculated from using
the bare vertex $\Gamma_{\gamma N\rightarrow\Delta}$ in the 
the upper half of Fig.\ 21. Similar to the results in Fig.\ 19 in the t-matrix
formulation, we see the large nonresonant mechanisms in dressing 
the $\gamma N \rightarrow \Delta$
vertex. The corresponding $E2/M1$ ratios $R_{EM}$ are compared in 
the lower half of Fig.\ 21. The nonresonant mechanisms change the ratio by
a factor of about 2 at resonance energy $W=1236$ MeV. 

In Table IV, we list the predicted $E_{1^+}$ and $M_{1^+}$ amplitudes of
the $\Delta\leftrightarrow \gamma N$ vertex evaluated at the
resonance energy $W=1236$ MeV.  
The parameters ($g_{\omega NN},G_M ,G_E$) = (10.5, 1.85, 0025) and (7., 1.95, -0.025)
from the fits to the data (Figs.\ 13 and 14) are used in this calculations.
We see that our average value $R_{EM}$ = (-1.8 $\pm$ 0.9)\% is
not too different from the average value
(-1.07 $\pm$ 0.37)\% of the empirical analysis\cite{davidson90}.
Since the assumption made in Ref.\ \cite{davidson90} is consistent with
our model as discussed above, the difference perhaps mainly comes from the
experimental uncertainties of the multipole amplitudes employed in the analysis of
Ref.\ \cite{davidson90}.  The differences between our predicted
multipole amplitudes and the empirical
values shown in Fig.\ 18 could also be responsible to this discrepancy. 
To compare our results with the values listed by the Particle Data Group
(PDG) \cite{particle}, we calculate the helicity amplitudes by
 
\begin{eqnarray}
A_{3/2}&=&\frac{\sqrt{3}}{2}[E_{1^+} - M_{1^+}]\,, \nonumber \\
A_{1/2}&=&-\frac{1}{2}[3 E_{1^+}+M_{1^+}] \,.\nonumber
\end{eqnarray}
The results at the resonance energy $W=1236$ MeV are listed in Table V.
The predictions from two constituent quark models\cite{bijker,capstick} are
also listed for comparision. We notice that our bare values are close to
the constituent quark model predictions\cite{capstick,bijker}, and 
the dressed values are close to the values
of PDG\cite{particle}. This suggest that our bare vertex can be identified with the
constituent quark model. The long-standing discrepancy between the
constituent quark model predictions and the PDG values is due to the
nonresonant meson-exchange production mechanisms which must be calculated
from a dynamical approach.
Similar considerations must be taken in comparing the PDG values with
the predictions of higher mass $N^*$ resonances from hadron models.

The results we have presented so far are based on the $\pi N$ model 
determined in a fit to the $\pi N$ phase shifts 
only up to 250 MeV. It is interesting to see the extent to which this model
can be extended to a higher energy region where the inelastic processes are still
not dominant. More importantly, we would like to examine whether the
extended model can yield significantly different $\pi N$ off-shell
dynamics which perhaps can help resolve 
the difficulty in reproducing 
the magnitudes of $\pi^+$ cross section (see Fig.\ 14b).
To  explore these possibilities, a $\pi N$ model is constructed by 
fitting the phase shifts data up to 400 MeV. 
This model is called Model-H to distingush it from Model-L 
from the fit up to only 250 MeV.
The resulting parameters are also listed in
Table 1. The phase shifts calculated from these two models
are compared in Fig.\ 22.
We see that Model-H (dotted curves) clearly gives a much better fit to the data in the
entire considered energy region. 
But it is not as accurate as Model-L in describing the
the crucial $P_{33}$ channel in the low energy region.
To accurately fit the $P_{33}$ in the
entire energy region and to resolve the difficulty in the $P_{13}$ channel, 
additional mechanisms may be needed.

The $\gamma N \rightarrow \pi N$ results calculated from using 
the Model-H and Model-L are compared in Fig.\ 23. 
The photon-asymmetry ratios (Figs.\ 23a,23c) are equally well described by both
models. They yield, however, significant differences in describing the
differential cross sections. In Fig.\ 23b, we see that Model-H gives 
a much better description of the $\pi^0$ differential cross sections 
in the high energy region. But it 
slightly overestimates the cross sections at low energies. The
$\pi^+$ differential
cross sections are better described by Model-H, as seen in Fig.\ 23d. 
But the difficulty in reproducing the magnitude is not removed entirely.
 
The results in Fig.\ 23 suggest that our predictions do depend to some
extent on the accuracy of the constructed $\pi N$ model in describing 
the $\pi N$ phase shifts. A natural next step is to extend
the present model to include the inelastic channels to obtain an accurate
fit up to 400 MeV. This extension then will introduce inelastic pion 
photoproduction
mechanisms, such as the $\gamma N \rightarrow \pi \Delta \rightarrow \pi N$
process, 
which may be needed to resolve the difficulty in getting an accurate
description of both the $\pi^0$ and $\pi^+$ processes. Such
a coupled-channel approach must also include the effect due to the
excitations of higher mass $N^*$ nucleon resonances. This must be pursued
in order to  make progress in using the 
forthcoming data from CEBAF to test hadron models.

\section{Conclusions and Future Studies}

We have applied the unitary transformation method first proposed in Ref.\ \cite{sato4}
to derive from a model Lagrangian with $N,\Delta ,\pi$ and $\rho$, $\omega$, and
$\gamma$ fields an effective Hamiltonian consisting of bare $\Delta \leftrightarrow \pi N$,
$\gamma N$ vertices and energy-independent meson-exchange $\pi N$ potential (Fig.\ 3)
and $\gamma N \rightarrow \pi N$ transition operator (Fig.\ 5.).  With the parameters
listed in Table 1 for the strong form factors and the bare mass of the $\Delta$, the
model can give a good description of $\pi N$ scattering phase shifts up to the $\Delta$
excitation energy region.  
The only adjustable parameters in the resulting 
pion photoproduction amplitude are the coupling strengths 
$G_E$ of the electric 
$E2$ and $G_M$ of the magnetic $M1$ transitions of the bare 
$\Delta \leftrightarrow \gamma N$ vertex and the less 
well-determined coupling constant $g_{\omega NN}$ of the $\omega$ meson.
We have shown that the best reproduction of the recent LEGS data of the 
photon-asymmetry ratios of the $\gamma p \rightarrow \pi ^0 p$ reaction depends
sensitively on these three parameters and yield $G_M = 1.9 \pm 0.05$ 
and $G_E = 0.0 \pm 0.025$ within the range $7 \leq g_{\omega NN}
\leq 10.5$. Within these ranges of parameters, the predicted differential cross
sections and photon-asymmetry ratios are in an overall good agreement with 
the data of $\gamma p \rightarrow \pi ^0 p$, $\gamma p \rightarrow \pi ^+ p$,
and $\gamma n\rightarrow \pi ^- p$ reactions from 180 MeV to the $\Delta$ excitation region.  
The model however underestimates the $\gamma N \rightarrow \pi N$ cross section at
energies above the $\Delta$ region.  This is expected since the constructed model
does not include inelastic channels, such as $\pi\Delta$, $\rho N$ channels, which
should start to play a significant role at energies above about 350 MeV.  Including
these channels could also be needed to resolve the difficulty in fitting $P_{13}$
$\pi N$ phase shifts (Fig.\ 22).
The constructed
effective Hamiltonian is free of the nucleon renormlization problem and hence is
suitable for nuclear many-body calculations.  

We have also analyzed the $K$-matrix method which is commonly used to extract
empirically the $\gamma N \rightarrow \Delta$ transition amplitudes from the 
$\gamma N \rightarrow \pi N$ data.  It is found that the
assumptions made in the $K$-matrix method \cite{davidson90} are consistent with
our meson-exchange dynamical model.  
Our average value of the $E2/M1$ ratio $R_{EM}$ = (-1.8 $\pm$ 0.9)\% is close to
(-1.07 $\pm$ 0.7)\% of Ref.\ \cite{davidson90}.
The helicity amplitudes
calculated from our bare $\gamma N \rightarrow \Delta$ vertex are in good agreement
with the predictions of the constituent quark models (Table IV).  The differences between these
bare amplitudes and the empirical values extracted from the data by using the $K$-matrix
method are shown to be due to the non-resonant meson-exchange mechanisms.  This suggests
that the bare vertex interactions in our effective Hamiltonian can be
identified with hadron models in which the $\pi N$ and $\pi\pi N$ ``reaction" channels
(both $\pi$ and $N$ are on their mass-shell) are excluded in the calculation
of the $N^*$ excitation.  Unfortunately we are not able to pin down the $E2/M1$ ratio
of the bare $\gamma \rightarrow \Delta$ vertex by considering the existing data
of photon-asymmetry ratios and differential cross sections.  More precise data of other
spin observables are needed to make progress.  This will be pursued when the data becomes
available, along with the extension of our approach to investigate pion electroproduction. 

The unitary transformation method developed here can be extended to higher
energy regions for investigating higher mass $N^*$ resonances.  To proceed, we need to
perform the unitary transformation up to second order in the 
coupling constants to account
for the 2$\pi$ production channels.  The resulting scattering equations will be
defined in a larger coupled channel space $N^* \oplus \pi N \oplus \gamma N \oplus \pi\pi N$.
This research program can be carried out in practice since the numerical methods for
solving such a Faddeev-type coupled-channel equations (because of the presence of the
three-body $\pi\pi N$ unitary cut) have been well developed \cite{aaron}.  Our effort
in this direction will be published elsewhere. 

\section*{Acknowledgements}
We would like to thank Andy Sandorfi for his very stimulating
discussions and providing experimental data.
One of the authors (T. Sato) would like to thank the
Theory Group of Physics Division at Argonne National Labolatory
for the hospitality and discussions.
This work is supported by the U. S. Department of
Energy, Nuclear Physics Division, under contract W-31-109-ENG-38,
and by Grant-in-Aid of Scientific Research, the Ministry of
Education, Science and Culture, Japan under contract 07640405.

\newpage

\appendix
\section*{Derivation of $\pi N$ potential}

To see how the Feynman-amplitude-like 
expressions of Eqs.\ (3.12)-(3.16) are obtained in our approach, we
give a detailed derivation of $\pi N$ potential from the following familiar Lagrangian

\begin{eqnarray}
{\cal{L}} &=& \Big[ - {f_{\pi NN} \over m_{\pi}}\, \bar{\psi} (x) \gamma_{5}
\gamma^{\mu} \vec{\tau}\, \psi (x) \cdot \partial_{\mu} \vec{\phi} (x) \\
\nonumber 
&+& {f_{\pi N\Delta} \over m_{\pi}} \vec\psi^{\mu}_{\Delta} (x) \roarrow{T}
\psi (x) \cdot \partial_{\mu} \vec{\phi} (x) \Big] + \Big[ h.c. \Big],
\end{eqnarray}
where $\psi (x)$, $\psi^{\mu}_{\Delta} (x)$, and $\phi (x)$ are
respectively the field operators for $N$, $\Delta$, and $\pi$; $\roarrow{T}$ is the
$\Delta \rightarrow N$ isospin transition operator, [h.c.] means taking the hermitian
conjugate of the first term.  By applying the 
canonical quantization procedure (see section III about the problem concerning
the $\Delta$ field), we can derive a Hamiltonian from Eq.\ (A1). 
The resulting Hamiltonian can be written as

\begin{equation}
H = H_0 + H_I 
\end{equation}
with

\begin{equation}
H_I = H^P_I + H^Q_I \, .
\end{equation}
where $H^P (H^Q )$ describes processes which can (cannot) take place in free space.
Explicitly, we can write in second quantization form 

\begin{eqnarray}
H^P_I &=& \sum_{\alpha} \int {1 \over (2\pi )^{3/2}} \, 
{1 \over \sqrt{2E_{\pi} (k)}} \, d\vec p \, d\vec p\,' \, d\vec k \\ 
\nonumber 
\Bigg\{& &i \, {f_{\pi N\Delta} \over m_{\pi}} \sqrt{m_N \over E_N (p)} \,
\sqrt{m_{\Delta} \over E_{\Delta} (p')} \, \bar{\omega}^{\mu}_{\vec p\,'} \,
T_{\alpha} \, u_{\vec p} \, k_{\mu} \, \Delta^{+}_{\vec p\,'} \,
b_{\vec p}\, a_{\vec k\,\alpha} \, \delta (\vec p + \vec k - \vec p\,' ) \\
\nonumber
& - & i \, {f_{\pi N\Delta} \over m_{\pi}} \, \sqrt{m_N \over E_N (p')} \,
\sqrt{m_{\Delta} \over E_{\Delta} (p)} \bar{u}_{\vec p\,'} \, T^{+}_{\alpha}\,
\omega^{\mu}_{\vec p} \, k_{\mu}\, b^{+}_{\vec p\,'} \, \Delta _{\vec p} \,
a^{+}_{\vec k , \alpha} \, \delta (\vec p - \vec k - \vec p\,' ) \Bigg\}
\end{eqnarray}
where $a^{+}_{k,\alpha}$, $b^{+}_{\vec p}$, and $\Delta ^{+}_{\vec p}$ are
, respectively, the creation operators for $\pi$, $N$, and $\Delta$ states, 
$\alpha$ is the pion isospin index, $u$ and $\omega^{\mu}$ are respectively the
spinors of Dirac and Rarita-Schwinger fileds.
Clearly, $H^P$ describes the $\Delta \leftrightarrow \pi N$ real processes 
(similar to Figs.\ 1a and 1b with the change $\sigma \rightarrow \Delta$)
which can take place in free space.  On the other hand, the virtual processes 
(similar to Figs.\ 1c-1f) are due to the following Hamiltonian

\begin{eqnarray}
H^Q_I &=& \sum_{\alpha} \int {1 \over (2\pi )^{3/2}} \, {1 \over \sqrt{2E_{\pi} (k)}} \,
d\vec p \, d\vec p\,' d\vec k \\
\nonumber
&\times & \Bigg[ i {f_{\pi NN} \over m_{\pi}}  \sqrt{m \over E_N (p)}  
\sqrt{m \over E_N (p')}  \Bigg\{ \bar u_{\vec p\,'}  \gamma _5  \tau^{\alpha}\not\! k 
u_{\vec p} b^{+}_{\vec p\,'} \, b_{\vec p} \Big[ -\delta (\vec p\,' - \vec p -
\vec k ) a_{\vec k,\alpha} + \delta (\vec p\,' - \vec p + \vec k) 
a^{+}_{\vec k,\alpha} \Big] \\
\nonumber
&+& \bar u_{\vec p\,'}\, \gamma _5 \, \tau^{\alpha} \not\! k \, v_{\vec p}\, b^{+}_{\vec p\,'}
d^{+}_{\vec p} \Big[ -\delta (\vec p\,' + \vec p - \vec k )\, a_{\vec k,\alpha} +
\delta (\vec p\,' + \vec p + \vec k )\, a^{+}_{\vec k,\alpha} \Big] \\
\nonumber
&+& \bar v_{\vec p\,'} \gamma _5 \not\! k \,\tau^{\alpha}\, u_{\vec p} \,d_{\vec p\,'}
b_{\vec p}\, \Big[ -\delta (-\vec p\,' - \vec p - \vec k )\, a_{\vec k,\alpha} + \delta
(-\vec p\,' - \vec p + \vec k ) a^{+}_{\vec k,\alpha} \Big] \\
\nonumber
&+& \bar v_{\vec p\,'} \gamma _5 \not\! k \tau^{\alpha} v_{\vec p}\, d_{\vec p\,'}\,
d^{+}_{\vec p}\, \Big[ - \delta (-\vec p\,' + \vec p - \vec k ) \, a_{\vec k,\alpha} +
\delta (-\vec p\,' + \vec p + \vec k ) \, a^{+}_{\vec k,\alpha} \Big] \Bigg\} \\
\nonumber
&-& i \, {f_{\pi N\Delta} \over m_{\pi}} \, \sqrt{m\over E_N (p')}\,
\sqrt{m_{\Delta} \over E_{\Delta}(p')} \Bigg\{
\bar{\omega}^{\mu}_{\vec p\,'} \, u_{\vec p}\, k_{\mu} \, \Delta^{+}_{\vec p\,'} \, b_{\vec p}\,
\delta (\vec p\,' - \vec p + \vec k ) a^{+}_{\vec k,\alpha} \\
\nonumber
&+& \bar{\omega}^{\mu}_{\vec p\,}\,\cdot v_{\vec p}\, k_{\mu} \,\Delta^{+}_{\vec p\,'}
d_{\vec p}^{+}\, \delta (\vec p\,' + \vec p + \vec k )\, a^{+}_{k,\alpha } \\
\nonumber
&+& i \,{f_{\pi N\Delta} \over m_{\pi}}\, \sqrt{m\over E_N (p')} \, 
\sqrt{m_{\Delta}\over E_{\Delta} (p)} \Bigg\{ \bar u_{\vec p\,'} \,\omega^{\mu}_{\vec p} \,
k_{\mu} \, b^{+}_{\vec p\,'} \,\Delta _{\vec p} \, \delta (\vec p\,' - \vec p - \vec k )
a_{\vec k,\alpha} \\ 
\nonumber
&+& \bar v_{\vec p\,'} \,\omega^{\mu}_{\vec p} \, k_{\mu} \,d_{\vec p\,'}\,
\Delta_{\vec p}\, \delta (-\vec p\,' - \vec p - \vec k ) a_{\vec k,\alpha} \Bigg\} 
\Bigg]
\end{eqnarray}
Note that the above equation includes an anti-nucleon spinor $v$, which is included
to maintain the relativistic feature of the starting quantum field theory. 

To proceed, we need to derive the unitary transformation operator
$S$. By the procedures outlined in section II, $S$ is related to $H^Q_I$. 
Hence, the actual task of deriving the $\pi N$ potential 
is to evaluate Eq.\ (2.23) from the $H^P_I$ and
$H^Q_I$ defined above. Let us first focus on
the first four $\pi NN$ coupling terms of Eq.\ (A.5).  
We need to consider 

\begin{eqnarray*}
|i> &=& b^{+}_{\vec p\,} \, a^{+}_{\vec k\alpha }\, | 0 > \\
|f> &=& b^{+}_{p\,'} \, a^{+}_{\vec k\alpha '} | 0 > 
\end{eqnarray*}
The allowed intermediate states in Eq.\ (2.23)
are $|n> = b^{+}_{\vec p_n} |0>$, $b^{+}_{\vec p_m}\, 
a^{+}_{k'\alpha '} \,a^{+}_{k\alpha}|0>$ for the first term, and
$|n> = d^{+}_{-\vec p_n}\,b^{+}_{\vec p\,'}\, b^{+}_{\vec p}\,
a^{+}_{\vec k\alpha} \, a^{+}_{\vec k\,'\alpha '} |0>$, $d^{+}_{-\vec p_m}
b^{+}_{\vec p\,'} b^{+}_{\vec p} |0>$ for the other three terms involving
the anti-nucleon component $v$.
Substituting these intermediate states into Eq.\ (2.23) and performing
straightforward operator algebra, we then obtain

\begin{eqnarray}
\langle \vec k\,' \alpha , \vec p\,'| H^{P}_{I} | \vec k \alpha, \vec p \rangle &=&
{(f_{\pi NN}/m_{\pi})^2 \over (2\pi )^{3}} \,
{1\over\sqrt{2E_{\pi}(k')}} \sqrt{m_N\over E_N (p')}   
\nonumber \\
&\times & \bar u_{\vec p\,'}\,
\Bigg[\sum^{4}_{i=1} M^{(i)} \Bigg] \sqrt{m_N \over E_N (p)} \, 
{1\over \sqrt{2E_{\pi} (k)}}\, u_{\vec p} \,,
\end{eqnarray}
where

\begin{eqnarray*}
M^{(1)} = {m_N \over E_N (p_n )}\, \gamma_5 \not\! k\,' \tau_{\alpha '}
u_{\vec p_n} \, \bar u_{\vec p\,'_n} \, \gamma_5 \not\! k\,\tau_{\alpha} {1\over 2}
&\Bigg[&{1\over E_N (p) - E_N (p_n ) + E_{\pi} (k)} \\
\\
&-& {1\over E_N (p_n) - E_N (p') - E_{\pi} (k')} \Bigg] \,,
\end{eqnarray*}
with $\vec p_n = \vec k + \vec p = \vec k' + \vec p\,'$;

\begin{eqnarray*}
M^{(2)} = {m_N\over E_N (p_m)}\,\gamma_5 \not\! k \,\tau_{\alpha} \,
u_{\vec p_m}\, \bar u_{\vec p_m} \gamma_5 \not\! k\,' \, \tau_{\alpha '}
{1\over 2} &\Bigg[& {1\over E_N (p) - E_N (p_m ) - E_{\pi} (k')}  \\
&-& {1\over E_N (p_m ) - E_N (p') + E_N (k)} \Bigg] \,,
\end{eqnarray*}
with $\vec p_m = \vec p - \vec k' = \vec p\,' - \vec k$;

\begin{eqnarray*}
M^{(3)} = - {m_N \over E_N (p_n )}\, \gamma_5 \not\! k\,' \,\tau_{\alpha '}\,
v_{-\vec p_n} \,\bar v_{-\vec p_n} \gamma_5 \not\! k \, \tau_{\alpha}
{1\over 2} &\Bigg[& {1\over -E_N (p') - E_N (p_n ) - E_{\pi} (k')} \\
&-& {1\over E_N (p) + E_N (p_n ) + E_{\pi} (k)} \Bigg] \,,
\end{eqnarray*}

\begin{eqnarray*}
M^{(4)} = - {m_N\over E_N (p_m )}\, \gamma_5 \not\! k \, \tau_{\alpha}
v_{-\vec p_m}\, \bar v_{-\vec p_m} \, \gamma_5 \not\! k\,'\, \tau_{\alpha '} {1\over 2}
&\Bigg[ &{1\over -E_N (p') - E_N (p_m ) + E_{\pi} (k)} \\
&-& {1\over E_N (p) + E_N (p_m ) - E_{\pi} (k')} \Bigg] \,,
\end{eqnarray*}
by using the properties that

\begin{eqnarray}
{m_N \over E_N (p)} \,u_{\vec p}\, \bar u_{\vec p} &=& {1\over 2E_N (p)} \,
\Big[ m_N + \gamma_{0}
E_N (p) - \vec{\gamma} \cdot \vec p \,\Big] \,,\\
\nonumber
{m_N \over E_N (p)} v_{\vec p} \bar v_{\vec p} &=& {1\over 2E_N (p)} \,\Big[
-m_N + \gamma _0 E_N (p) - \vec{\gamma} \cdot \vec p \,\Big] \,,
\end{eqnarray}
one can easily show that for an arbitrary $p_0$

\begin{eqnarray}
&&{m_N \over E_N (p_n )} u_{\vec p_n} \bar u_{\vec p_n} \,{1\over p_0 - E_N (p_N )} +
{m_N \over E_N (p_N )}\, v_{-\vec p_n} \bar v_{-\vec p_n} \,{1\over p_0 + E_N (p_N )} \\
\nonumber
&=& {1\over 2E_N (p_n )} \Bigg[ (m_N - \vec{\gamma} \cdot \vec p_n ) 
\Bigg(\,{1\over p_0 - E_N (p_n )} - {1\over p_0 - E_N (p_n )}\, \Bigg) \\
\nonumber
&+& \gamma _0 E_N (p_n ) \Bigg(\, {1\over p_0 - E_N (p_n )} +
{1\over p_0 + E_N (p_n )} \,\Bigg) \Bigg] \\
\nonumber
&=& {1\over p^{2}_{0} - E^{2}_{N} (p_n )} \Big[ m_N- \vec{\gamma} \cdot
\vec p _n + \gamma _0 p_0 \Big] = {\not\! p _n + m_N \over p^{2}_{n} - m_N^2 } =
{1\over \not\! p _n - m_N }
\end{eqnarray}
where $p_n = (p_0 \,, \vec p_n )$.

By using Eq.\ (A8), we can combine various propagators in Eq.\ (A6) to obtain

\begin{eqnarray*}
\sum^{4}_{i=1} M^{(i)} &=& \gamma _5 \not\! k\,' \, \tau _{\alpha '} \,{1\over 2}\,
\Big[ {1\over (\not\! p \,+ \not\! k ) - m_N} + 
{1\over (\not\! p +  \not\! k\,') -m_N} \Big]
\gamma _5 \not\! k \, \tau_{\alpha} \\
&+& \gamma _5 \not\! k \,\tau_{\alpha} \,{1\over 2} \,
\Big[{1\over (\not\! p - \not\! k\,') -m_N} 
+ {1\over (\not\! p\,' - \not\!  k ) -m_N } \Big]
\gamma _5 \,\not\! k\,' \,\tau_{\alpha '}
\end{eqnarray*}
where $p = (E_N (p) , \vec p)$, $k = (E_{\pi} (k) , \vec k)$.  The above result looks
remarkably simple.  It resembles very much the usual Feynman amplitudes, except
that the intermediate nucleon propagator is the average of two Dirac propagators
for the momenta evaluated using the incoming or outgoing $\pi N$ momentum variables.

The evaluation of the $\Delta$ terms is much more involved, but yields a similar
form as given in Eqs.\ (3.15)-(3.16).
Similar derivations can also be carried out to define the $\pi N$ interactions 
, Eq.\ (3.14), due to the $\rho$ meson.

\newpage

\begin{table}
\caption{The parameters of the $\pi N$ models. The units are
1/Fermi for cutoff parameters $\Lambda_{\alpha}$ and MeV for the bare $\Delta$ mass.
Model-L and Model-H are obtained respectively 
from fitting $\pi N$ phase shifts up to 250 MeV and
400 MeV.}
\begin{tabular}{ldddddddd}
Model & ${f^{2}_{\pi NN} \over 4\pi}$ & $\Lambda_{\pi NN}$ & $g_{\rho NN} g_{\rho\pi\pi}$ &
$\kappa _{\rho}$ & $\Lambda _{\rho}$ & $f_{\pi N\Delta}$ & $\Lambda _{\pi N \Delta}$ &
$m _{\Delta}$ \\
\hline
Model-L & 0.08 & 3.2551 & 38.4329 & 1.825 & 6.2305 & 2.049 & 3.29 & 1299.07 \\
Model-H & 0.08 & 3.7447 & 39.0499 & 2.2176 & 7.5569 & 2.115 & 3.381 & 1318.52 \\ 
\multicolumn{9}{c}{Form factor $F_{\pi NN}(k)= \left( {\Lambda ^{2}_{\pi NN} \over 
\Lambda^{2}_{\pi NN}+k^2} \right) ^2$ \, , \, 
$F_{\rho} (q) = \left( {\Lambda^{2}_{\rho} \over \Lambda^{2}_{\rho} + \vec{q}\,^2}
\right) ^2$ \, , \,
$F_{\pi N\Delta}(k) = \left( {\Lambda^{2}_{\Lambda N\Delta} 
\over \Lambda ^{2}_{\pi N\Delta} + k^2 }\right) ^2$ } \\
\end{tabular}
\end{table}
\bigskip
\bigskip
\bigskip
\begin{table}
\caption{The calculated $\pi N$ scattering lengths( in unit of Fermi) are
compared with the values determined in Ref.\ \protect\cite{koch}}
\begin{tabular}{lccc}
& Model-L & Model-H & Koch-Pietarinen \\
\hline
$S_{11}$ &  0.1588 &  0.1737 &  0.173 $\pm$ 0.003 \\
$S_{31}$ & -0.1191 & -0.1198 & -0.101 $\pm$ 0.004 \\
$P_{11}$ & -0.0976 & -0.0864 & -0.081 $\pm$ 0.002 \\
$P_{31}$ & -0.0509 & -0.0478 & -0.045 $\pm$ 0.002 \\
$P_{13}$ & -0.0363 & -0.0383 & -0.030 $\pm$ 0.002 \\
$P_{33}$ &  0.2523 &  0.2797 &  0.214 $\pm$ 0.002 \\
\end{tabular}
\end{table}

\newpage
\begin{table}
\caption{The parameters for the $\gamma N \rightarrow \pi N$ interactions defined
in Eqs.\ (4.16)-(4.20) and Fig.\ 5.}
\begin{tabular}{lcc}
$f_{\pi NN}, \Lambda_{\pi NN}$ & -- & Table I \\
$f_{\pi N\Delta}, \Lambda_{\pi N\Delta} $ & -- & Table I \\
$g_{\rho NN} = \sqrt{g_{\rho NN} g_{\rho\pi\pi}}$ & -- & Table I \\
$\Lambda_{\rho NN}=\Lambda_{\rho}, \kappa_{\rho}$  & -- & Table I \\
$g_{\rho\pi\gamma}$  &  0.1027  &  Ref.[45]  \\
$g_{\omega\pi\gamma}$ & 0.3247 & Ref.[45] \\
$g_{\omega NN}$ & 7 -- 10.5  &  See text \\
$\Lambda_{\omega NN} = \Lambda_{\rho}$ &  --  & Table I \\
$\kappa_{\omega} =0$ &  --  & See text \\
$G_M (0)$ & 1.85 -- 2.0  &  See text \\
$G_E (0)$ & $\pm$ 0.025 & See text \\
\end{tabular}
\end{table}
\bigskip
\bigskip
\bigskip
\newpage
\begin{table}
\caption{The magnetic $M_{1^+}$ and electric $E_{1^+}$ 
amplitudes of $\Delta \rightarrow \gamma N$ transition
at $W = 1236$ MeV. $R_{EM} = E_{1^+}/M_{1^+}$.
The amplitudes are in unit of $10^{-3}$ (GeV)$^{-1/2}$.
The numbers in the upper and lower rows for each
case are respectively from using ($g_{\omega NN},G_M , G_E$) =
(10.5, 1.85, + 0.025) and (7., 1.95, -0.025).}
\begin{tabular}{ccccc}
 &  $M_{1^+}$  &  $E_{1^+}$ & $R_{EM}$ & Average \\
\hline
$\Gamma_{\Delta \rightarrow \gamma N}$ (Bare)  &  175  &  -2.28  & -1.3\% \\
&&&&  (0.0 $\pm$ 1.3)\%  \\
   &  184  &  +2.28  &  +1.2\%  \\
\\
$\bar{\Gamma}_{\Delta \rightarrow\gamma N}$ (Dressed)  &  257  & -6.9 & --2.7\% \\
&&&&  (-1.8 $\pm$ 0.9)\% \\
  &  258  &  -2.26  &  -0.9\% 
\end{tabular}
\end{table}
\bigskip
\bigskip
\begin{table}
\caption{Helicity amplitudes of the $\Delta \rightarrow \gamma N$ transition
at $W = 1236$ MeV are compared with the values from Particle Data Group (PDG) [45]
and the predictions of constituent quark models of Refs.[47,48].  The amplitudes
are in unit of $10^{-3}$ (GeV)$^{-1/2}$.
The numbers in the upper and lower rows for each
case are respectively from using ($g_{\omega NN},G_M , G_E$) =
(10.5, 1.85, + 0.025) and (7., 1.95, -0.025).}
\begin{tabular}{lccccc}
 A&  PDG & Dressed & Bare & Ref.\ [48] & Ref.\ [47] \\
\hline
$A_{3/2}$& -257 $\pm$ 8  & -228 & -153 & -157 & -186 \\
  &  &  -225  &  -158 \\
$A_{1/2}$& -141 $\pm$ 5   & -118 &  -84 & -91  & -108  \\
  &  &  -126  &  -96
\end{tabular}
\end{table}
\bigskip
\bigskip
\bigskip
\newpage

\begin{figure}
\caption{Graphical representation of the interaction Hamiltonians $H^P_I$ of
Eq.\ (2.7) and $H^Q_I$ of Eq.\ (2.8).}
\end{figure}

\begin{figure}
\caption{Graphical representation of the effective interaction Hamiltonians
$H^{'P}_I$ of Eq.\ (2.19) and $H^{'Q}_I$ of Eq.\ (2.20).}
\end{figure}

\begin{figure}
\caption{Graphical representation of the interactions of the effective
Hamiltonian Eq.\ (3.8) in the coupled $\pi N\oplus \Delta$ space.}
\end{figure}

\begin{figure}
\caption{Graphical representation of scattering equations defined by 
Eqs.\ (3.23)-(3.30).}
\end{figure}

\begin{figure}
\caption{Graphical representation of the effective interactions 
$\Gamma_{\Delta \leftrightarrow \gamma N}$ and $v_{\gamma N}$ of Eq.\ (4.13).}
\end{figure}

\begin{figure}
\caption{Graphical representation of pion production amplitudes defined
by Eqs.\ (4.22)-(4.24).}
\end{figure}

\begin{figure}
\caption{The $\pi N$ phase shifts calculated from the $\pi N$ Model-L of
Table I are compared with the empirical values of the analyses of
Ref.\ \protect\cite{hohler} (open squares) and Ref.\ \protect\cite{arndt} (solid squares).
$T_L$ is the pion laboratory energy.}
\end{figure}

\begin{figure}
\caption{The on-shell matrix elements of $\pi N$ potentials 
defined by Eqs.\ (3.11)-(3.16). $T_L$ is the pion laboratory energy.
The notations are $N_D:v_{N_D}$, $N_E:v_{N_E}$, $\rho:v_\rho$,
$\Delta_D:v_{\Delta_D}$, $\Delta_E:v_{\Delta_E}$, and TOT is the sum.}
\end{figure}

\begin{figure}
\caption{The upper half shows the 
mass of the $\Delta$ state defined by Eq.\ (3.30). $m_{\Delta}=1299.07$ MeV is the bare mass
of Model-L of Table I, $M_R=1236$ is the experimental resonance position, and
$W$ is the $\pi N$ invariant mass.  The lower half shows the 
bare and dressed $\Delta \rightarrow \pi N$ form factors defined by Eqs.\ (5.1)
and (5.2) with $\bar{f}_{\pi N \Delta}=1.3 f_{\pi N\Delta}$. $k$ is the pion
momentum in the $\Delta$ rest frame.}
\end{figure}

\begin{figure}
\caption{The $\pi N$ phase shifts ($\delta$) and scattering amplitude 
($t(k_0,k_0,W)$) with the $\pi N$ invariant mass
$W = E_N (k_0 ) + E_{\pi} (k_0 )$ in the $P_{11}$ channel.
The solid curves are from the full calculation. The dashed curves are from 
the calculation including only the 
nucleon pole potential $v_{N_D}$ of Fig.\ (3c).}
\end{figure}

\begin{figure}
\caption{The photon-asymmetry ratios 
$R_\gamma=d\sigma_{\parallel}/d\sigma_{\perp}$ of
$\gamma p \rightarrow \pi^0 p$ at $90^0$ degrees.
$E_{\gamma}$ is the photon energy in the laboratory frame.  The results in the
upper half are from using $G_M=1.85, G_E=0.025$ and three values of
$g_{\omega NN}$. The results in the lower half are from using 
$G_M=1.85, g_{\omega NN}=10.5$ and four values of $G_E$.
The data are from Ref.\ \protect\cite{legs,sandorfi}.}
\end{figure}

\begin{figure}
\caption{The region of the parameters $(g_{\omega NN}, G_M, G_E)$ for
describing the data of $\gamma N \rightarrow \pi N$ reactions.  See text for the
explanations.}
\end{figure}

\begin{figure}
\caption{The photon-asymmetry ratios(a) and differential cross sections(b) 
for $\gamma p \rightarrow \pi^0 p$ reaction
at four angles in the center of mass frame. The solid and dotted
curves are respectively fom the calculation using the 
parameters $(g_{\omega NN}, G_M,G_E)=(10.5,1.85,+0.025)$, and
$(7.,1.95,-0.025)$. The data are from Refs.\ \protect\cite{legs,sandorfi}.
$E_{\gamma}$ is the photon energy in the laboratory frame.}
\end{figure}

\begin{figure}
\caption{Same as Fig.\ 13, except for the $\gamma p \rightarrow \pi^+ n$ reaction.
The data are from Ref.\ \protect\cite{sandorfi}.}
\end{figure}

\begin{figure}
\caption{The photon-asymmetry ratios(a) and differential cross sections(b)
for the $\gamma p \rightarrow \pi^0 p$ reaction at four angles.
The solid and dotted curves are respectively from the calculations using the
parameters $(g_{\omega NN}, G_M,G_E)=(10.5,2.0,+0.025)$, and $(7., 2.10,-0.025)$.
The data are from Ref.\ \protect\cite{legs,sandorfi}.
$E_{\gamma}$ is the photon energy in the laboratory frame.}
\end{figure}

\begin{figure}
\caption{Same as Fig.\ 15, except for the $\gamma p \rightarrow \pi^+ n$ reaction.
The data are from Ref.\ \protect\cite{sandorfi}.}
\end{figure}

\begin{figure}
\caption{Differential cross sections of $\gamma p \rightarrow \pi^0 p$, (a)
$\gamma p \rightarrow \pi^+ n$, (b) and $\gamma n \rightarrow \pi^- p$ (c)
reactions. The solid (dotted) curves are calculated by using
$G_M=1.85(2.0)$. Both calculations using the same 
($g_{\omega NN}, G_E )=(10.5,+0.025)$. The data are from 
Ref.\protect\cite{menze}.
$E_{\gamma}$ is the photon energy in the laboratory frame.}

\end{figure}

\begin{figure}
\caption{The predicted multipole amplitudes $M_{1^+}$ and $E_{1^+}$ in 
the total isospin $I=3/2$ channel are
compared with the empirical values of Refs.\ \protect\cite{arndt} (solid squares) and
\protect\cite{berends} (open squares). The parameters used in this calculation are
$G_M=1.85$, $g_{\omega NN}=10.5$, with $G_E=0.025$ (solid curve)
and  -0.025 (dotted curve).
$E_{\gamma}$ is the photon energy in the laboratory frame.}
\end{figure}

\begin{figure}
\caption{The $M_{1^+}$ and $E_{1^+}$ multipole amplitudes of the dressed vertex 
$\bar{\Gamma}_{\gamma N \rightarrow \Delta}$ defined by Eq.\ (4.24) and the bare
vertex $\Gamma_{\gamma N \rightarrow \Delta}$ are compared. The dressed vertex is 
a complex function written as $\Gamma(\alpha)=|\Gamma(\alpha)|e^{i\phi(\alpha)}$
with $\alpha = M_{1^+}, E_{1^+}$.  $W$ is the $\gamma N$ invariant mass.}
\end{figure}

\begin{figure}
\caption{The predicted K-matrix defined by Eq.\ (4.28). The solid curves
are the full calculations. The dotted curves are from the 
resonant term. The dashed curves (denoted as $B$) are the contributions from the
nonresonant term $k_{\gamma \pi}$ defined by Eq.\ (4.29).
$W$ is the $\gamma N$ invariant mass.}
\end{figure}

\begin{figure}
\caption{The $M_{1^+}$ residues A of the $K$-matrix (Eq.\ (4.32))
calculated from the dressed 
$\Delta \rightarrow \gamma N$ defined by Eq.\ (4.30) and the bare
vertex are compared in the upper half. 
Their corresponding ratios
$R_{EM}=\frac{E_{1^+}}{M_{1^+}}$ are compared in the lower half.
$W$ is the $\gamma N$ invariant mass.}
\end{figure}

\begin{figure}
\caption{The $\pi N$ phase shifts calculated from Model-L (solid curves)
and Model-H (dotted curves) are compared. The data are 
from Refs.\ \protect\cite{arndt} (solid squares)
and \protect\cite{hohler} (open squares).  $T_L$ is pion energy in the
laboratory frame.}
\end{figure}

\begin{figure}
\caption{Photon-asymmetry ratios and the differential cross sections for
the $\gamma p \rightarrow \pi^0 p$ (a and b) and $\gamma p \rightarrow
\pi^+n$ (c and d) reactions.
The solid (dotted) curves are from calculations using $\pi N$ Model-L (Model-H).
The parameters are $(g_{\omega NN}, G_M,G_E)=(10.5,1.85,+0.025)$. The
data are from Ref.\ \protect\cite{legs,sandorfi}.
$E_{\gamma}$ is the photon energy in laboratory frame.}
\end{figure}

\end{document}